\documentclass[12pt]  {revtex4}
\usepackage{graphicx}
\usepackage{mathrsfs}
\usepackage{mdwlist}
\usepackage{subfigure}
\usepackage{booktabs}
\usepackage{amsmath}
\usepackage{cases}
\usepackage{extarrows}
\usepackage{dsfont}
\usepackage{amstext}
\usepackage{amsbsy}
\usepackage{bbm}
\usepackage{bm}
\usepackage{amssymb,mathrsfs,amsmath}
\usepackage{threeparttable}
\usepackage{booktabs}
\usepackage{bbding,pifont}
\usepackage{amsthm}
\usepackage{graphicx}
\usepackage{textcomp}
\usepackage{multirow}%
\usepackage{pifont}
\usepackage{color}
\usepackage{sansmath}
\usepackage{makecell}
\usepackage{tabularx}
\usepackage[colorlinks,citecolor=blue]{hyperref}
\definecolor{Dgreen}{RGB}{0, 100, 0}
\usepackage{url}
\usepackage[colorlinks]{hyperref}
\hypersetup{%
	plainpages=true,
	breaklinks=true,  
	hypertexnames=false,  
	pageanchor=true,
	colorlinks=true,
	linkcolor={blue},
	citecolor={blue},
	urlcolor={blue},
	anchorcolor={black}
}

\begin{document}

\title{Three-state coherent control using narrowband and passband sequences}

\author{Cheng Zhang}
\affiliation{Fujian Key Laboratory of Quantum Information and Quantum Optics, Fuzhou University, Fuzhou
350108, China}
\affiliation{Department of Physics, Fuzhou University, Fuzhou 350108, China}

\author{Li-Tuo Shen}
\affiliation{Fujian Key Laboratory of Quantum Information and Quantum Optics, Fuzhou University, Fuzhou
350108, China}
\affiliation{Department of Physics, Fuzhou University, Fuzhou 350108, China}

\author{Jie Song}
\affiliation{Department of Physics, Harbin Institute of Technology, Harbin 150001, China}

\author{Yan Xia}
\affiliation{Fujian Key Laboratory of Quantum Information and Quantum Optics, Fuzhou University, Fuzhou
350108, China}
\affiliation{Department of Physics, Fuzhou University, Fuzhou 350108, China}

\author{Zhi-Cheng Shi}\thanks{szc20147@163.com}
\affiliation{Fujian Key Laboratory of Quantum Information and Quantum Optics, Fuzhou University, Fuzhou
350108, China}
\affiliation{Department of Physics, Fuzhou University, Fuzhou 350108, China}

\begin{abstract}
In this work,
we propose a comprehensive design for narrowband and passband composite pulse sequences by involving the dynamics of all states in the three-state system.
The design is quite universal as all pulse parameters can be freely employed to modify the coefficients of error terms.
Two modulation techniques,
the strength and phase modulations,
are used to achieve arbitrary population transfer with a desired excitation profile,
while the system keeps minimal leakage to the third state.
Furthermore,
the current sequences are capable of tolerating inaccurate waveforms, detunings errors,
and work well when rotating wave approximation is not strictly justified.
Therefore,
this work provides versatile adaptability for shaping various excitation profiles in both narrowband and passband sequences.
\end{abstract}

\maketitle

\section{Introduction}

In recent decades,
the composite pulse (CP) technique \cite{Levitt1979,WIMPERIS199046,WIMPERIS1989509,TYCKO198590,Levitt1986} has gained widespread attraction in quantum information processing \cite{PhysRevApplied.18.034062,PhysRevLett.129.240505,PhysRevA.108.032614}.
This technique was originally developed in fields of polarization optics \cite{Peters:12} and nuclear magnetic resonance (NMR) \cite{COUNSELL1985133,Wimperis1991,JONES201191,ODEDRA201268,ODEDRA201281}.
Since then, it has been utilized in various branches of physics \cite{PhysRevA.70.052318,Cummins_2000,
PhysRevA.80.032303,PhysRevA.90.012316,PhysRevA.90.033608,PhysRevA.93.023830,PhysRevA.87.043418,
PhysRevA.93.043419,PhysRevApplied.13.014048,PhysRevResearch.4.023222,PhysRevA.106.033102} with a particular focus on quantum control \cite{PhysRevA.103.032609,
PhysRevApplied.10.054051,
PhysRevA.101.012321,
PhysRevA.104.012609,PhysRevA.80.024302,Shi2022,Xu22}.
The CP sequence,
a series of constant pulses with appropriate relative phases,
has reliably accomplished several desired quantum tasks and efficiently compensated errors caused by external noises \cite{RevModPhys.76.1037,PhysRevA.87.052317,PhysRevLett.106.233001,PhysRevLett.113.043001,PhysRevA.89.022341}.
A well-known one is the broadband sequence \cite{PhysRevA.67.012317,PhysRevA.99.013402,PhysRevA.107.023103,PhysRevA.106.042613,PhysRevA.105.042414},
which is widely adopted to improve the error-resistant ability for the implementation of single- and multi-qubit quantum gates \cite{PhysRevA.67.042308,PhysRevA.92.022333,PhysRevA.84.062311,PhysRevA.93.032340,
PhysRevA.90.012341,PhysRevA.92.060301,PhysRevResearch.2.043194,
PhysRevA.103.052612,PhysRevA.73.032334}.
However,
it is not a panacea and sometimes unsuitable for some specific tasks,
such as addressing operations in trapped ions \cite{HAFFNER2008155} and optical lattice systems \cite{PhysRevA.74.042344,PhysRevA.75.053612,PhysRevLett.99.020502}.
On these occasions,
the narrowband (NB) sequence \cite{TYCKO1984462,Ivanov2011,SHAKA1984169,PhysRevA.102.013105,PhysRevA.84.065404,PhysRevA.107.032618} offers a more suitable candidate.

The NB sequence aims at enhancing sensitivity to systematic errors.
More specifically,
system states are allowed to evolve only within a specific error threshold,
and their evolutions are inhibited when falling below this threshold \cite{PhysRevA.84.065404}.
In essence,
the system states remain unchanged once the parameter deviates significantly from the precise value \cite{PhysRevA.83.053420}.
This unique feature makes the NB sequence perfectly adaptive to local addressing operations in quantum computation \cite{Ivanov2011},
because the excitation of the particles surrounding the object can be efficiently inhibited.
Additionally,
in order to enhance the resilience of this addressing operation against minor systematic errors,
the passband (PB) sequence was developed \cite{PhysRevA.88.063410,PhysRevLett.56.1905,PhysRevA.90.040301,PhysRevA.83.053420,PhysRevA.84.063413}.
The versatility of the PB sequence is demonstrated in two primary ways:
it can successfully correct small errors in parameters,
and significantly prohibit the system evolution when the errors are large enough \cite{PhysRevA.88.063410}.
Recently,
the NB sequence has proven to be a valuable aid in spatial localization for biological or medical NMR spectroscopy \cite{TYCKO1984462},
while the PB sequence has emerged as an efficient tool for selecting high-intensity target signals and suppressing unwanted background signals in NMR experiments \cite{Husain2013}.

So far,
the researches on NB and PB sequences \cite{PhysRevA.107.032618,PhysRevA.83.053420,PhysRevA.84.065404,
Ivanov2011,SHAKA1984169,PhysRevA.102.013105,
PhysRevA.88.063410,PhysRevLett.56.1905,PhysRevA.90.040301,TYCKO1984462} have primarily focused on two-state systems.
When performing  individual addressing operations on a three-state system,
the sequences derived from two-state systems  cannot be directly applied due to their different structures.
Moreover, there is only one coupling strength in the two-state system while it has  two coupling strengths in the three-state system.
This trait allows one to adjust the ratio of the two coupling strengths to design broadband sequences in the three-state system \cite{PhysRevA.105.042414}.
Another significant difference is the participation of the third state in the system evolution.
In a two-state system,
only the population of one state needs to be calculated,
because the other state is naturally obtained according to the normalization condition.
In contrast,
if there is leakage to the third state, population transfer would be incomplete,
leading to imperfect quantum control in a three-level system.

In previous work \cite{PhysRevResearch.2.043194},
with the help of the Morris-Shore transformation \cite{PhysRevA.27.906} and Majorana decomposition \cite{Majorana1932},
the three-state system can be reduced into an effective two-state one.
As a result,
it is impossible to consider the dynamics of the excited state, completely neglecting population leakage to the excited state.
In the presence of leakage,
the well-designed CP sequences \cite{PhysRevA.83.053420} may be invalid in three-state systems.
Hence,
it is necessary to contain the dynamics of all system states for NB and PB sequences.
On the other hand,
predetermining physical parameters (except for the modulation parameters) may fail to fully nullify derivatives,
ultimately jeopardizing the construction of NB and PB sequences in the three-state system.
Therefore,
a general method needs to be developed for constructing NB and PB sequences applied to various types of modulations in three-state systems.

In this paper,
we construct the NB and PB sequences for arbitrary population transfer in the three-state system.
The design procedure entails modifying the error coefficients in the transition probability,
thereby identifying appropriate modulation parameters for the NB and PB sequences.
When the error coefficients cannot be completely nullified,
we propose a cost function to obtain a suitable substituted solution that minimizes the coefficients as much as possible.
Two modulation techniques, the strength and phase modulations,
are employed for creating the NB sequences with almost identical shapes of the excitation profile,
as well as the PB sequences with a desired excitation profile.
Furthermore,
both NB and PB sequences exhibit significant reduction in leakage to the third state.
The numerical results indicate that the strength modulation has adaptability to inaccurate waveforms,
while the phase modulation shows resistance to detuning errors.
Besides,
both modulations show resistance to the case that the rotating wave approximation (RWA) is not strictly justified.

This paper is organized as follows.
In Sec.~\ref{model},
we introduce the physical model and propose the general design method for NB and PB sequences in the three-state system.
In Sec.~\ref{pulse design},
we illustrate how to construct the NB and PB sequences with specific pulse numbers by the strength and phase modulations.
For each modulation,
the sequences up to seven pulses are investigated.
In Sec.~\ref{arbitrary},
we take five pulses to exemplify the construction of the NB and PB sequences for arbitrary population transfer between two lower states.
In Sec.~\ref{environment},
we demonstrate the validity of the current sequences when the pulse is imperfect.
The conclusion is given in Sec.~\ref{con}.

\section{Physical model and general theory}\label{model}

\begin{figure*}[b]
\centering
\scalebox{1}{\includegraphics{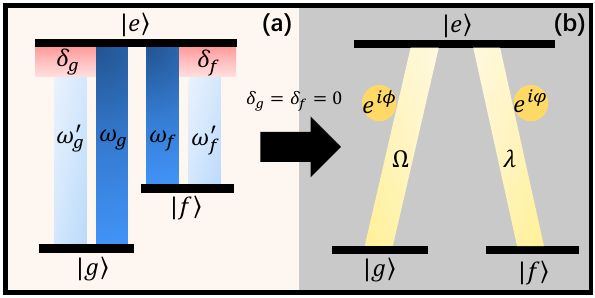}}
\caption{A $\Lambda$-type three-state system driven by two control fields
(a) in the original picture and (b) in the rotating frame.}\label{mod}
\end{figure*}

Let us consider a paradigmatic three-state system with the $\Lambda$-type structure,
as shown in Fig.~\ref{mod}(a).
The transition frequency between the lower state $|g\rangle (|f\rangle)$ and the excited state $|e\rangle$ is $\omega_{g(f)}=\left[E_e-E_{g(f)}\right]/\hbar$ with level energy $E_k (k=g,f,e)$.
Here,
two lower states are forbidden for direct transition and coupled to the excited state by two control fields $\Omega(t)=\Omega\cos{(\omega'_g t+\phi)}$ and $\lambda(t)=\lambda\cos{(\omega'_f t+\varphi)}$ with frequencies $\omega_{g}'$ and $\omega_{f}'$,
coupling strengths $\Omega$ and $\lambda$, and phases $\phi$ and $\varphi$,
respectively.
The Hamiltonian of the system reads ($\hbar=1$ hereafter)
\begin{eqnarray}\label{norwa}
H=\sum_{k=g,f,e}E_k|k\rangle \langle k| +\Omega  \cos{(\omega'_g t+\phi)}|g\rangle\langle e|+\lambda  \cos{(\omega'_f t+\varphi)}|f\rangle\langle e|+\mathrm{H.c.}
\end{eqnarray}
In the rotating frame $R=|g\rangle \langle g|+\exp{(i\omega'_ft)}|f\rangle \langle f|+\exp{(i\omega'_gt)}|e\rangle \langle e|$,
the Hamiltonian under RWA can be rewritten as
\begin{eqnarray}\label{Hami}
H=\frac{\Omega}{2}e^{i \phi}|g\rangle\langle e|+\frac{\lambda}{2}e^{i \varphi}|f\rangle\langle e|+\mathrm{H.c.},
\end{eqnarray}
where we assume $\delta_{g}=\omega_{g}-\omega_{g}'=0$ and $\delta_{f}=\omega_{f}-\omega_{f}'=0$ for simplicity,
as shown in Fig.~\ref{mod}(b).
Note that the RWA is well satisfied in the atomic system driven by laser fields,
where the coupling strength $\Omega(\lambda)$ can be varied via manually controlling the intensity of laser fields,
while the phase $\phi$($\varphi)$ can be regulated by adjusting the phase of laser fields \cite{PhysRevLett.118.083604,Wollenhaupt2005,PhysRevLett.101.246809,PhysRevA.100.043403} via an electro-optical or acousto-optical modulator \cite{PhysRevA.88.063410}.

When the Hamiltonian (\ref{Hami}) is time-independent,
the propagator at the evolution time $T$ reads $U(T)=\exp(-iHT)$.
It is instructive to adopt its matrix form in the basis $\{|g\rangle, |f\rangle, |e\rangle\}$ up to a global phase
\renewcommand\arraystretch{0.3}
\begin{eqnarray}\label{propagator}
U   =    \left[\begin{array}{ccc} 	
   \cos^2 {\theta} + \cos {\frac{A}{2}}  \sin^2 {\theta}
&    \sin^2 {\frac{A}{4}}\sin {2\theta}e^{i (\phi-\varphi)}
&    -i \sin {\frac{A}{2}}\sin {\theta}e^{i \phi}\\[2ex]
   \sin^2 {\frac{A}{4}}\sin {2\theta}e^{i (\varphi-\phi)}
&   \sin^2 {\theta} + \cos {\frac{A}{2}} \cos^2 {\theta}
&    -i \sin {\frac{A}{2}}\cos {\theta}e^{i \varphi} \\[2ex]
  -i\sin {\frac{A}{2}}\sin {\theta}e^{-i \phi}
&    -i\sin {\frac{A}{2}}\cos {\theta}e^{-i \varphi}
&    \cos {\frac{A}{2}}   \cr
\end{array}	
\right] ,
\end{eqnarray}
where $A=\Omega_0T$ is the pulse area.
We define
$\Omega=\Omega_0\sin{\theta}$ and $\lambda=\Omega_0\cos{\theta}$,
and thus $\tan{\theta}=\Omega/\lambda$,
where $\theta$ is called the coupling strength ratio.

The propagator of the $N$-pulse sequence is yielded by a product of the propagators of single resonant pulses,
i.e.,
\begin{eqnarray}\label{UN}
U^{(N)}(NT)=U_N U_{N-1} \cdots U_2U_1,
\end{eqnarray}
where $U_n$ represent the evolution operator of the $n^{\mathrm{th}}$ pulse,
$n=1,2,\cdots,N$.
Obviously, the total propagator depends on all modulation parameters of the ingredient pulses,
labeled as $\{A_n,\theta_n,\phi_n,\varphi_n\}$.

Assume that each pulse has identical pulse area $2\pi$ and the same pulse area error $\epsilon$,
i.e.,
the actual pulse area for each pulse reads $A_n=2\pi(1+\epsilon)$.
According to the definition of the pulse area,
the error $\epsilon$ refers to either the coupling strength error $\Omega_0(1+\epsilon)$ or the pulse duration error $T(1+\epsilon)$ under the resonance condition.
In this way,
the transition probability of the state $|f\rangle$ can be labelled as $P^{(N)}_f(\bm{\theta},\bm{\phi},\bm{\varphi},\epsilon)$,
where $\bm{\theta}=(\theta_1,\cdots,\theta_N)$,
$\bm{\phi}=(\phi_1,\cdots,\phi_N)$ and $\bm{\varphi}=(\varphi_1,\cdots,\varphi_N)$.
The primary objective is to implement population transfer between the states $|g\rangle$ and $|f\rangle$ in a desired manner.
Namely,
the transition probability of the target state satisfies
\begin{eqnarray}\label{populationinversion}
P^{(N)}_f(\bm{\theta},\bm{\phi},\bm{\varphi},0)=\mathcal{P},
\end{eqnarray}
where $\mathcal{P}$ is a predetermined value.
Remarkably,
complete population inversion is achieved by setting $\mathcal{P}=1$.

Generally speaking,
for the NB sequence,
the transition probability must drop sharply around the target pulse area $A_n=2\pi$ and be ``frozen'' around $A_n=4\pi$ in the excitation profile.
The physical effect of this setting is that quantum operations are extremely sensitive to errors,
so that one can achieve precise addressing control in dispersion external fields.
Mathematically,
the excitation profile for the NB sequence is acquired via nullifying the derivatives of $P^{(N)}_f(\bm{\theta},\bm{\phi},\bm{\varphi},\epsilon)$ at $\epsilon=\pm1$ in order.
We can seek the modulation parameters of the NB sequence,
i.e.,
$\bm{\theta}$, $\bm{\phi}$ and $\bm{\varphi}$,
through fulfilling a group of equations on the derivatives
\begin{eqnarray}\label{eqNB}
\tilde{x}_{N,m}(\bm{\theta},\bm{\phi},\bm{\varphi})=\frac{1}{m!}\frac{\partial^{m}}{\partial\epsilon^{m}}\left[P^{(N)}_f(\bm{\theta},\bm{\phi},\bm{\varphi},\epsilon)\right]\Big|_{\epsilon=\pm1}=0,
\end{eqnarray}
where $\tilde{x}_{N,m}(\bm{\theta},\bm{\phi},\bm{\varphi})$ is the coefficient of the $m^{\mathrm{th}}$ order error term in the Taylor expansion at $\epsilon=\pm1$,
and $m=4,6,\cdots,M_N$.
Note that $\tilde{x}_{N,2}$ and all odd-order coefficients naturally vanish due to the choose of $A_n=2\pi$.

For the PB sequence,
the transition probability needs to maintain unchanged around $A_n=2\pi$ meanwhile vanish around $A_n=4\pi$.
To this end,
we have to nullify the error terms in the transition probability at both $\epsilon=0$ and $\epsilon=\pm1$,
and thus the modulation parameters are obtained by solving the following equations
\begin{numcases}{}\label{eqPB}
~\tilde{x}_{N,m}(\bm{\theta},\bm{\phi},\bm{\varphi})=\frac{1}{m!}\frac{\partial^{m}}{\partial\epsilon^{m}}\left[P^{(N)}_f(\bm{\theta},\bm{\phi},\bm{\varphi},\epsilon)\right]\Big|_{\epsilon=\pm1}=0, \nonumber\\[-1.5ex]
~~~\\[-1.5ex]
x_{N,m'}(\bm{\theta},\bm{\phi},\bm{\varphi})=\frac{1}{m'!}\frac{\partial^{m'}}{\partial\epsilon^{m'}}\left[P^{(N)}_f(\bm{\theta},\bm{\phi},\bm{\varphi},\epsilon)\right]\Big|_{\epsilon=0}=0,\nonumber
\end{numcases}
where $x_{N,m'}(\bm{\theta},\bm{\phi},\bm{\varphi})$ is the coefficient of the ${m'}^{\mathrm{th}}$ order error term in the Taylor expansion at $\epsilon=0$,
$m=4,6,\cdots,M_P$,
and $m'=2,4,\cdots,M'_P$.
From the view of the physical picture, the excitation profile exhibits narrow wings with a large $M_P$,
while its central flat top becomes wider as $M'_P$ increases.
Without causing ambiguity,
we omit the symbol $(\bm{\theta},\bm{\phi},\bm{\varphi})$ in expressions hereafter.

It is worth noting that the above design method may not succeed,
since either Eqs.~(\ref{eqNB}) or (\ref{eqPB}) sometimes are unsolvable.
On this occasion,
we can identify a substituted group of solutions to significantly reduce low-order error coefficients.
The substituted solution can be obtained by minimizing a cost function established by Eqs.~(\ref{eqPB}),
which has different forms for the NB and PB sequences:
\begin{subequations}
\begin{eqnarray}
\mathrm{Narrowband}:\mathcal{F}_{\mathrm{NB}}^{(N)}&\!\!\!\!\!=\!\!\!\!\!&\sum_{m}^{M_N} c_{m}|\tilde{x}_{N,m}|,\label{filter1}\\[-0.8ex]
\mathrm{Passband}:\mathcal{F}_{\mathrm{PB}}^{(N)}&\!\!\!\!\!=\!\!\!\!\!&\sum_{m}^{M_P} c_{m}|\tilde{x}_{N,m}|+\sum_{m'}^{M'_P} c_m' |x_{N,m'}|,~~~~~~\label{filter2}
\end{eqnarray}
\end{subequations}
where $c_{m(m')}$ is the weight factor.
To ensure that the cost function gives more weight to low-order error coefficients and alleviate the contribution of high-order ones,
the weight factor should decrease monotonously with an increase in $m(m')$.

Alternatively,
by setting $0<\mathcal{P}<1$,
the current method is also feasible to construct the NB and PB sequences for arbitrary population transfer between two lower states in the three-state system.
Unlike two-state systems \cite{PhysRevLett.110.140502,PhysRevB.95.241307},
we have to consider the evolution of the third state,
and the leakage to the third state must be suppressed to restrict the system dynamics within two states $\{|g\rangle, |f\rangle\}$.
Hence,
apart from satisfying Eqs.~(\ref{populationinversion})-(\ref{eqPB}),
the modulation parameters of NB and PB sequences should meet the following equations
\begin{numcases}{}\label{leakeq}
\tilde{y}_{N,l}=\frac{1}{l!}\frac{\partial^l}{\partial\epsilon^l}\left[P^{(N)}_e(\epsilon)\right]\Big|_{\epsilon=\pm1}=0,\nonumber\\[-1.5ex]
\\[-1.5ex]
y_{N,l'}\!=\frac{1}{l'!}\frac{\partial^{l'}}{\partial\epsilon^{l'}}\left[P^{(N)}_e(\epsilon)\right]\Big|_{\epsilon=0}=0,\nonumber
\end{numcases}
where $P^{(N)}_e(\epsilon)$ is the transition probability of the third state $|e\rangle$,
$l=2,4,\cdots,L$, and $l'=2,4,\cdots,L'$.
Again, $y_{N,0}$, $\tilde{y}_{N,0}$, and all odd-order coefficients in Eq.~(\ref{leakeq}) vanish by setting $A_n=2\pi$.
Next,
we elucidate the CP design in detail by exemplifying specific pulse numbers.

\section{Pulse sequence design}\label{pulse design}
In this section,
we demonstrate how to design various kinds of CP sequences in the three-state system.
Two modulation techniques,
the strength and phase modulations,
are adopted in the design procedure.
For a certain modulation,
all parameters are fixed except for the modulation one.
\subsection{Strength modulation}\label{str}
We first elaborate on the construction of NB and PB sequences by the  strength modulation.
For brevity,
the $N$-pulse NB and PB sequences by the strength modulation are called as the S-NB$N$ and S-PB$N$ sequences,
respectively.
The phase differences in each pulse are held on constant,
e.g.,
$\phi_n=\pi/2$ and $\varphi_n=0$ with $n=1,2,\cdots,N$.
Noting that there are two coupling strengths $\Omega_n$ and $\lambda_n$ for the $n^{\mathrm{th}}$ pulse,
we label the coupling strength ratio as $\tan{\theta_n}=\Omega_n/\lambda_n$,
and $\theta_n$ are recognized as the modulation  parameters.
Physically, we can modulate one of two coupling strengths while fixing another to obtain the variables $\theta_n$.

The target propagator of the $N$-pulse sequence in the absence of errors is
\renewcommand\arraystretch{0.5}
\begin{eqnarray}\label{UNN}
U^{(N)}=\left[
\begin{array}{ccc}	
\cos{\tilde{\theta}_N}&-i\sin{\tilde{\theta}_N}&0  \\[.3ex]
i^{(2N+3)}\sin{\tilde{\theta}_N}&(-1)^N\cos{\tilde{\theta}_N}&0  \\[.3ex]
0&0&(-1)^N  \\[.3ex]
\end{array}	
\right],~~
\end{eqnarray}
where $\tilde{\theta}_N=2\sum^{N}_{n=1}(-1)^{n+1}\theta_n$.
Obviously,
to ensure complete population inversion,
the strength ratio of the first pulse can be chosen as
\begin{eqnarray}\label{theta1}
\theta_1=\sum_{n=2}^N (-1)^{n}\theta_n+\frac{(2k+1)\pi}{4},
\end{eqnarray}
where $k$ is an arbitrary integer.
Through this selection,
we have
\begin{eqnarray}
\tilde{\theta}_N=\frac{\pi}{2}+2k\pi.
\end{eqnarray}
Then,
the remaining $(N-1)$ ratios are obtained by solving Eqs.~(\ref{eqNB}) for the S-NB$N$ sequence or Eqs.~(\ref{eqPB}) for the S-PB$N$ sequence.

\begin{figure}[t]
\centering
\scalebox{0.45}{\includegraphics{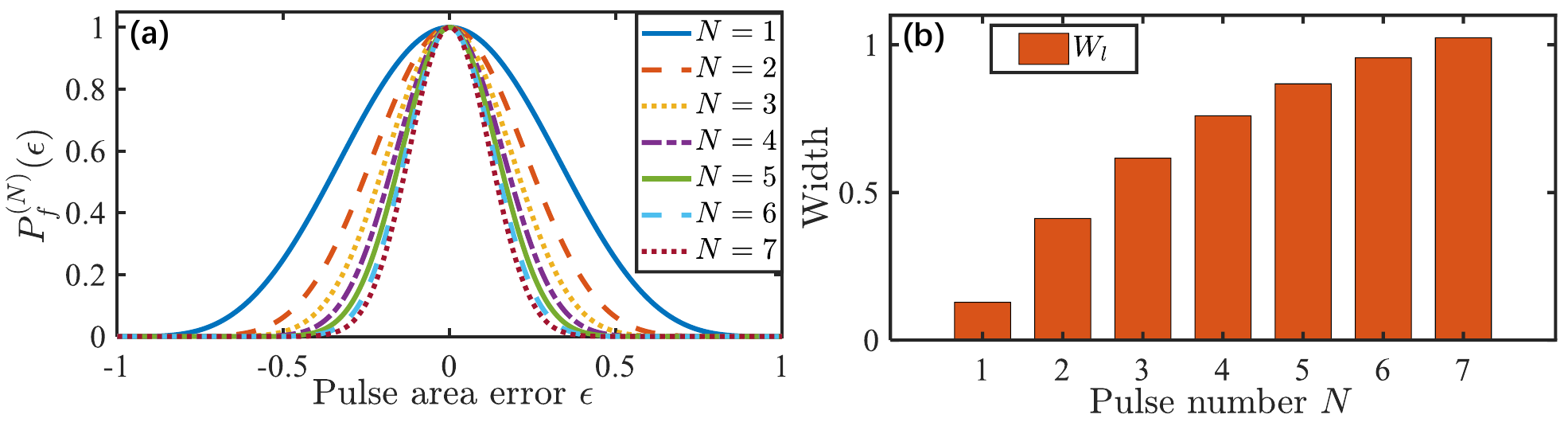}}
\caption{(a) Excitation profiles for the S-NB$N$ sequences.
(b) Low excitation width $W_l$ vs the pulse number $N$ in the S-NB$N$ sequences.
For $N=1$,
$\theta_1=\pi/4$ and $\phi_1=\varphi_1=0$.
All $\theta_n$ come from Table~\ref{tab1}.}\label{twopfig}
\end{figure}

For examples,
to construct the S-NB$2$ sequence for complete population inversion between $|g\rangle\leftrightarrow|f\rangle$,
we need to solve Eqs.~(\ref{populationinversion}) and (\ref{eqNB}) with $M_N=4$,
i.e.,
\begin{subequations}
\begin{eqnarray}
x_{2,0}&\!\!\!\!=\!\!\!\!&\sin^2{2(\theta_1-\theta_2)}=1,\\[-0.4ex]
\tilde{x}_{2,4}&\!\!\!\!=\!\!\!\!&\frac{\pi^4}{4}\big[\sin{\theta_1}\!\cos{\theta_1}\!+\!\cos{\theta_2}(2\sin{\theta_1}\!+\!\sin{\theta_2})\big]^2=0,
\end{eqnarray}
\end{subequations}
and one group solution for $\theta_1$ and $\theta_2$ are ($k=0$)
\begin{eqnarray}
\theta_1=\theta_2+\frac{\pi}{4},~~~~\theta_2=\frac{5\pi}{8}-\frac{1}{2}\arctan{\sqrt{2(1+\sqrt{2})}}.
\end{eqnarray}
As for the design of S-NB3 sequence,
all $\theta_n$ are derived in a similar way,
through setting $M_N=6$,
and the numerical solutions are given in Table~\ref{tab1}.
We can continue to combine more pulses for longer NB sequences.
Figure~\ref{twopfig}(a) displays the excitation profiles for the single pulse and the S-NB$N$ sequences from $N=2$ to $N=7$, respectively.
Apparently,
with the pulse number $N$ increasing,
the excitation profile becomes narrower.

\renewcommand\arraystretch{1.2}
\begin{table*}[t]
\caption{Coupling strength ratios $\theta_n$ for the S-$\mathrm{NB}N$ sequences (in units of $\pi$).\label{tab1}}
\setlength\tabcolsep{7.2pt}
 \centering
 \begin{tabular}{ccccccccccccccc}
 \hline
  \hline
    &$\theta_1$&$\theta_2$&$\theta_3$&$\theta_4$&$\theta_5$&$\theta_6$&$\theta_7$\\
  \hline
  S-NB2&0.6930&0.4430&--&--&--&--&--\\[-1.5ex]
  S-NB3&0.9223&0.9908&0.3185&--&--&--&--\\[-1.5ex]
  S-NB4&1.8111&0.3617&1.7659&0.9653&--&--&--\\[-1.5ex]
  S-NB5&0.8578&0.3304&1.4755&1.3296&1.5767&--&--\\[-1.5ex]
  S-NB6&1.1688&1.7614&0.5511&0.3724&1.5905&0.9266&--\\[-1.5ex]
  S-NB7&0.7487&1.9199&1.2087&1.5952&0.3258&0.8483&0.3301\\[0.3ex]
  \hline
  \hline
 \end{tabular}
\end{table*}

In order to quantitatively measure the magnitude of the efficient excitation region,
we define the low excitation width $W_l$ to represent the bilateral low excitation region,
i.e.,
\begin{eqnarray}
W_l=W_{\mathrm{right}}+W_{\mathrm{left}}=|1-\epsilon^+_{l}|+|1+\epsilon^-_{l}|,
\end{eqnarray}
where $W_{\mathrm{right}(\mathrm{left})}$ denotes the width of the low excitation region on right (left) side,
and $\epsilon^{+}_l$ and $\epsilon^{-}_l$ are two solutions for the transition probability satisfying $P^{(N)}_f(\epsilon)=10^{-4}$.
As shown in Fig.~\ref{twopfig}(b),
$W_l$ gradually enlarges with the increasing of the pulse number.
In particular,
the S-NB$N$ sequence with few pulses (e.g., $N=2$ or $N=3$) can significantly compress the excitation region,
leading to a remarkable improvement of the sensitivity around $\epsilon=0$.
For the S-NB7 sequence,
the excitation profile has a wide range of the low excitation on both sides ($W_l=1.022$).

\begin{figure}[t]
\centering
\scalebox{0.8}{\includegraphics{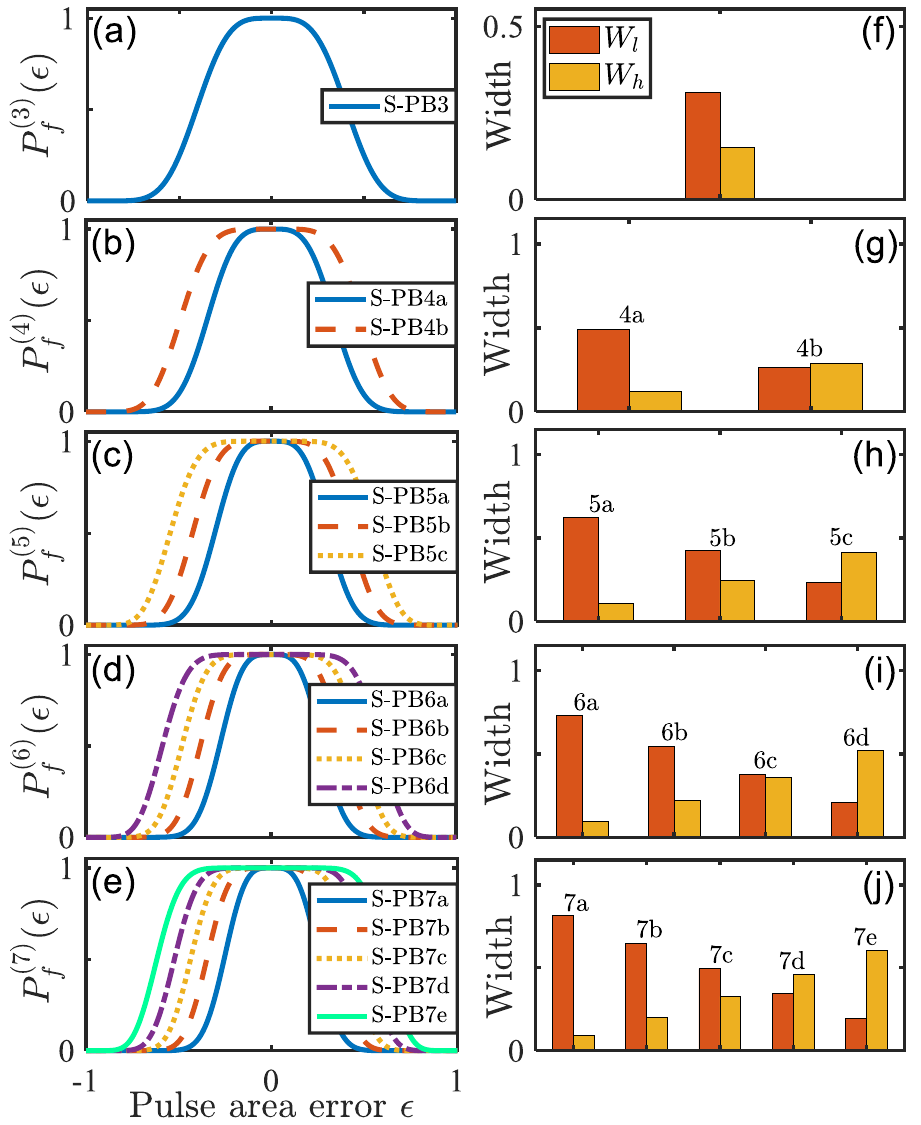}}
\caption{Excitation profiles (left column) and the corresponding widths $W_l$ and $W_h$ (right column) for the S-PB3, S-PB4X, S-PB5X, S-PB6X and  S-PB7X sequences.
All $\theta_n$ are presented in Table~\ref{tab2}.
}\label{fig5}
\end{figure}

Note that
we are unable to obtain the excitation profile for the S-PB2 sequence,
because three subequations in Eqs.~(\ref{populationinversion}) and (\ref{eqPB}) cannot be simultaneously satisfied by only two modulation parameters.
To construct the S-PB3 sequence,
we set $M_P=4$ and  $M'_P=2$ in Eqs.~(\ref{eqPB}),
and the solutions are given in Table~\ref{tab2}.
As depicted in Figs.~\ref{fig5}(a) and \ref{fig5}(f),
the excitation profile for the S-PB3 sequence displays a wide flat top in the central region,
while the width of the low excitation region becomes moderately small.
This means that the obtained S-PB3 sequence not only makes selective excitation within a localized pulse area region,
but also achieves population inversion with robustness against small pulse area error.

\renewcommand\arraystretch{1.2}
\begin{table*}[t]
\centering
\caption{Coupling strength ratios for the S-PB$N$X sequences (in units of $\pi$).\label{tab2}}
\setlength\tabcolsep{9.5pt}
 \begin{tabular}{cccccccccccccccc}
 \hline
  \hline
&$\theta_1$&$\theta_2$&$\theta_3$&$\theta_4$&$\theta_5$&$\theta_6$&$\theta_7$\\[0.5ex]
  \hline\\[-3.5ex]
\!\!\!S-PB3&0.5537&1.4910&1.7227&--&--&--&--\\[-1.5ex]
  S-PB4a&1.9842&0.0475&0.0238&1.7104&--&--&--\\[-1.5ex]
 S-PB4b&0.9036&0.6474&0.4615&0.4677&--&--&--\\[-1.5ex]
  S-PB5a&0.1409&1.9082&1.5695&0.4220&0.8698&--&--\\[-1.5ex]
  S-PB5b&0.0835&0.0770&1.5016&1.1888&1.9306&--&--\\[-1.5ex]
S-PB5c&1.7192&0.2221&0.8691&0.0492&1.9330&--&--\\[-1.5ex]
  S-PB6a&0.2738&0.4828&1.3861&1.4450&1.8460&1.3281&--\\[-1.5ex]
  S-PB6b&1.9064&1.9160&0.3679&0.0166&1.1768&1.2686&--\\[-1.5ex]
  S-PB6c&1.9305&1.6977&1.4651&1.4409&1.5025&1.5095&--\\[-1.5ex]
  S-PB6d&0.1531&0.1932&1.3768&1.5882&0.3105&1.8089&--\\[-1.5ex]
  S-PB7a&0.9832&1.0867&0.7896&1.3600&1.1954&1.7292&1.4578\\[-1.5ex]
  S-PB7b&0.5937&1.1840&0.6068&0.2113&0.4956&0.5423&0.4916\\[-1.5ex]
  S-PB7c&0.9369&1.1978&0.9863&1.3774&0.5201&0.3942&0.7759\\[-1.5ex]
  S-PB7d&0.3728&1.6973&0.1163&1.4978&0.9549&1.6558&1.6569\\[-1.5ex]
  S-PB7e&1.9388&1.7566&0.2566&0.8841&0.1531&1.7732&0.3155\\
  \hline
  \hline
 \end{tabular}
\end{table*}

As long as the pulse number exceeds three,
we can design various shapes for the excitation profile in S-PB$N$ sequences,
i.e.,
different widths of the flat top and two low excitation bottoms.
Here, based on the shape of the excitation profile, we further subdivide and relabel them as the S-PB$N$X sequences,
where $\mathrm{X}=\mathrm{a}, \mathrm{b}, \mathrm{c},\cdots$ corresponds to $M'_P=4,6,8,\cdots,2N-4$ and $M_P=2N-M'_P$ in Eqs.~(\ref{eqPB}).
Figures~\ref{fig5}(b)-(e) show the excitation profiles for the S-PB$N$X sequences from $N=4$ to $N=7$.
To quantify the robustness against the small error,
we define the width of the high excitation region [$P^{(N)}_f(\epsilon)\geq0.999$],
which reads
\begin{eqnarray}
W_h=|\epsilon^+_h-\epsilon^-_h|,
\end{eqnarray}
where $\epsilon^{+}_h$ and $\epsilon^{-}_h$ are two solutions satisfying $P^{(N)}_f(\epsilon)=0.999$.
In Figs.~\ref{fig5}(g)-(j),
$W_l$ and $W_h$ are displayed for different types of S-PB$N$X sequences.
For a certain type, e.g., the type `a',
with the pulse number $N$ increasing,
the low excitation width gradually enlarges in their excitation profiles.
For the same pulse number,
different types of seqeunces have different $W_l$ and $W_h$.
Therefore, by solving different subequations in Eqs.~(\ref{eqPB}),
we can make the system only excited within a specific pulse area region.

\subsection{Phase modulation}\label{pha}

Next,
we construct the NB and PB sequences by using the phase modulation in which $\phi_n$ and $\varphi_n$ are recognized as the modulation parameters.
For simplicity,
the coupling strength ratio is fixed as $\theta_n=\pi/4$,
i.e.,
$\Omega_n=\lambda_n$.
In this situation,
the CP sequence must be combined by odd pulses, because the system would return to the initial state when pulses are even.
Besides,
the phases of the first pulse can be arbitrary, since they do not affect population transfer \cite{PhysRevA.99.013402}.
Thus,
we simply set $\phi_1=\varphi_1=0$.
Similar to the strength modulation,
we term the NB sequence with $N$ pulses as the P-NB$N$ sequence.

\begin{figure}[t]
\centering
\scalebox{0.6}{\includegraphics{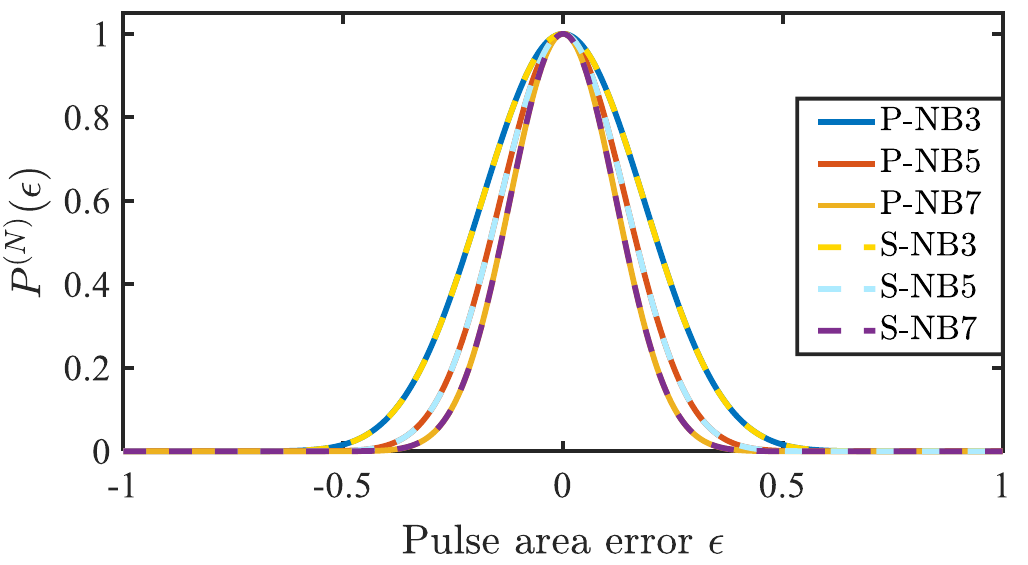}}
\caption{Excitation profiles for the P-NB$N$ and S-NB$N$ sequences.
The modulation parameters can be found in either Table~\ref{tab1} or \ref{tab3}.
}\label{FIG.6}
\end{figure}

\renewcommand\arraystretch{1.2}
\begin{table*}[t]
\centering
\caption{Phases for the P-NB$N$ sequences (in units of $\pi$).\label{tab3}}
\setlength\tabcolsep{20.2pt}
 \begin{tabular}{ccccccccccccccccccccccccc}
 \hline
  \hline
~&P-NB3&P-NB5&P-NB7\\
  \hline\\[-5.5ex]
$\phi_2$&1.0000&0.8890&1.2301\\[-1.5ex]
$\varphi_2$&1.3333&1.2884&0.7066\\[-1.5ex]
$\phi_3$&1.6667&1.0475&0.7358\\[-1.5ex]
$\varphi_3$&0.3333&0.0787&1.1614\\[-1.5ex]
$\phi_4$&--&0.2278&1.4494\\[-1.5ex]
$\varphi_4$&--&0.9377&0.3906\\[-1.5ex]
$\phi_5$&--&1.6684&0.0099\\[-1.5ex]
$\varphi_5$&--&1.2184&1.6341\\[-1.5ex]
$\phi_6$&--&--&0.8954\\[-1.5ex]
$\varphi_6$&--&--&1.3972\\[-1.5ex]
$\phi_7$&--&--&1.6944\\[-1.5ex]
$\varphi_7$&--&--&0.4215\\[0.3ex]
  \hline
  \hline
 \end{tabular}
\end{table*}

The target propagator of the $N$-pulse sequence in the absence of errors reads (up to a global phase)
\renewcommand\arraystretch{0.5}
\begin{eqnarray}\label{pUNN}
U^{(N)}
=\left[
\begin{array}{ccc}	
0&e^{-i \sum^N_{n=1}(-1)^n\Psi_n}&0  \\[2ex]
e^{i \sum^N_{n=1}(-1)^n\Psi_n}&0&0 \\[2ex]
0&0&1 \\
\end{array}	
\right],
\end{eqnarray}
where $\Psi_n=\phi_n-\varphi_n$ represents the phase difference in each pulse.
We readily find from Eq.~(\ref{pUNN}) that the transition probability of the target state $|f\rangle$ automatically reaches unity.
Therefore,
the total $2(N-1)$ phases for the P-NB$N$ sequence can be obtained by solving Eqs.~(\ref{eqNB}) with $M_N=4N-2$,
and the numerical solutions are given in Table~\ref{tab3}.
Figure~\ref{FIG.6} shows the excitation profiles for the pulse number $N=3, 5$ and $7$.
For a comparison,
the P-NB$N$ sequences are also plotted in Fig.~\ref{FIG.6}.
We can see that the P-NB$N$ and S-NB$N$ sequences create almost identical excitation profile under the same pulse number.
The reason can be found as follows.
In the phase modulation,
the transition probability is accurate up to the $(4N-2)^{\mathrm{th}}$ order for the P-NB$N$ sequence.
On the other hand,
by the S-NB$N$ sequence,
all coefficients from $\tilde{x}_{N,2N+2}$ to $\tilde{x}_{N,4N-2}$ are extremely low and can be approximately regarded as nullification.
As a result,
both P-NB$N$ and S-NB$N$ sequences can almost eliminate the same number of error terms in Eqs.~(\ref{eqNB}) and (\ref{eqPB}).

\begin{figure}[t]
\centering
\scalebox{0.6}{\includegraphics{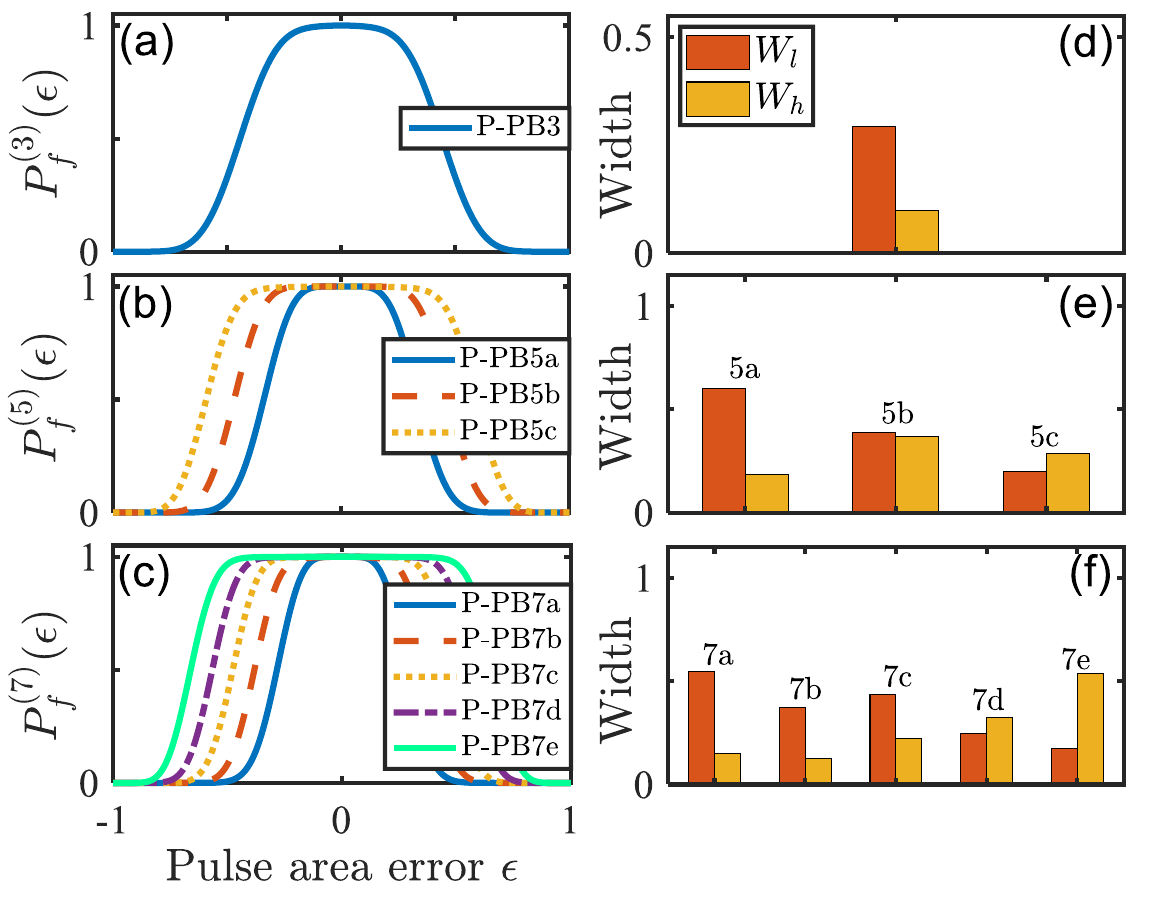}}
\caption{Excitation profiles (left column) and the corresponding widths $W_l$ and $W_h$ (right column)
for the P-PB3,
 P-PB5X and  P-PB7X sequences.
All $\phi_n$ and $\varphi_n$ are presented in Table~\ref{tab4}.
}\label{FIG.7}
\end{figure}

It is worth mentioning that there are no solutions for the P-PB3 sequence by solving Eqs.~(\ref{eqPB}) with $M_P=6$ and  $M'_P=4$.
According to Eq.~(\ref{filter2}),
we can find a substituted solution by minimizing the following cost function
\begin{eqnarray}\label{filter3}
\mathcal{F}^{(3)}_{\mathrm{PB}}=e^{-2}|x_{3,2}|+e^{-4}(|x_{3,4}|+|\tilde{x}_{3,4}|)+e^{-6}|\tilde{x}_{3,6}|
\end{eqnarray}
with the corresponding weight factor $e^{-m}$, and the numerical solution of phases for the P-PB3 sequence are presented in Table~\ref{tab4}.
It is observed in Fig.~\ref{FIG.7}(a) that even though the substituted solution of the phases cannot fulfill Eqs.~(\ref{eqPB}),
the excitation profile for the P-PB3 sequence still exhibits a desired central flat top and bilateral flat bottoms.
Moreover,
the high and low excitation widths are similar to those in the excitation profile for the S-PB3 sequences.

\renewcommand\arraystretch{1.2}
\begin{table*}[t]
\centering
\caption{Phases for the P-$\mathrm{PB}N\mathrm{X}$ sequences (in units of $\pi$).\label{tab4}}
\setlength\tabcolsep{5.8pt}
 \begin{tabular}{lccccccccccccccccccccccccc}
 \hline
  \hline
  ~&$\mathrm{PB}3$&$\mathrm{PB}5\mathrm{a}$&$\mathrm{PB}5\mathrm{b}$&$\mathrm{PB}5\mathrm{c}$&$\mathrm{PB}7\mathrm{a}$&$\mathrm{PB}7\mathrm{b}$&$\mathrm{PB}7\mathrm{c}$&$\mathrm{PB}7\mathrm{d}$&$\mathrm{PB}7\mathrm{e}$\\
  \hline\\[-5.5ex]
$\phi_2$&1.6631&1.3226&1.1066&0.5545&0.8918&1.0043&0.7689&0.0801&0.4857\\[-1.5ex]
$\varphi_2$&1.3708&1.9758&1.4412&0.6957&1.3931&1.4798&1.4526&0.6923&1.0137\\[-1.5ex]
$\phi_3$&1.0588& 0.1176&0.5884 &1.2167&1.5376&1.5550&0.7495 &0.1371&0.1068\\[-1.5ex]
$\varphi_3$&0.3485&0.9756&0.0506&1.6382&0.2099&1.0034&1.8600&1.9693&1.1928\\[-1.5ex]
$\phi_4$&--&0.9860&1.0003&1.7003&0.5670&1.1009&1.3410&0.5222&1.9721 \\[-1.5ex]
$\varphi_4$&--&1.0843&1.9541&1.1905&0.4894&1.9481&1.0660&0.9819&0.2282\\[-1.5ex]
$\phi_5$&--&1.5012&0.2322&0.7557&1.2734&0.2563&1.5433&0.0705&0.7782\\[-1.5ex]
$\varphi_5$&--&1.9778&1.1589&1.5723&1.5472&1.5372 &1.4520&1.5411&1.3012\\[-1.5ex]
$\phi_6$&--&--&--&--&0.1126&1.8322&1.2466&1.5615&0.7032\\[-1.5ex]
$\varphi_6$&--&--&--&--&1.1991&0.2665&1.2015&1.9327&1.9502\\[-1.5ex]
$\phi_7$&--&--&--&--&0.7005&1.9778&0.4884&0.5160&1.3763\\[-1.5ex]
$\varphi_7$&--&--&--&--&1.6726&1.1120&0.2337&0.7303&0.7019\\[0.3ex]
  \hline
  \hline
 \end{tabular}
\end{table*}

When the pulse number $N>3$,
there are also various types of P-PB$N$ sequences.
Similarly,
according to the shape of the excitation profiles,
we further call them the P-PB$N$X sequences, where $\mathrm{X}=\mathrm{a},\mathrm{b},\mathrm{c},\cdots$ correspond to $M'_P=4,8,12\cdots,4N-8$ and $M_P=4N-2-M'_P$.
We present the substituted solutions for the P-PB5X and P-PB7X sequences in Table~\ref{tab4},
and display the excitation profiles in Figs.~\ref{FIG.7}(b) and \ref{FIG.7}(c), respectively.
Compared to the S-PB$N$X sequence,
the P-PB$N$X sequences also exhibit similar but not identical robustness and selectivity; see the slightly different $W_l$ and $W_h$ in Figs.~\ref{fig5}(h),~\ref{fig5}(j),~\ref{FIG.7}(e) and \ref{FIG.7}(f).
The reason is that,
although the error coefficients are not entirely nullified in the P-PB$N$X sequences,
they are suppressed to an extremely low level instead.
Therefore,
both modulation techniques are capable of obtaining the PB sequence with a desired excitation profile.

\section{Arbitrary population transfer between two lower states}\label{arbitrary}

\begin{figure*}[t]
\centering
\scalebox{0.8}{\includegraphics{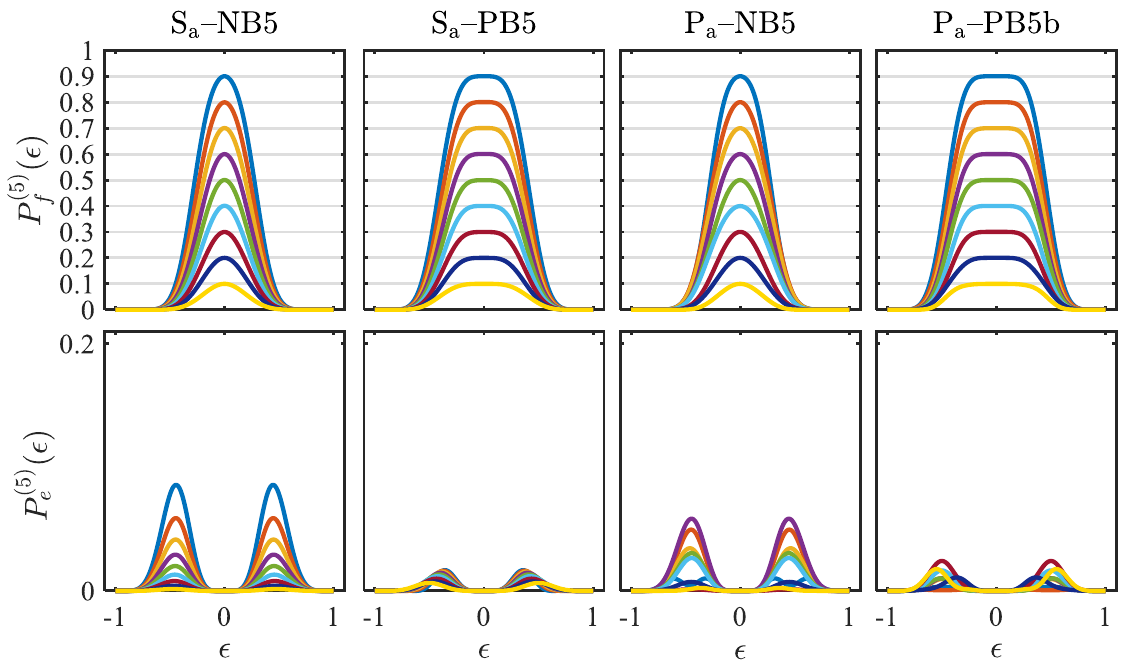}}
\caption{Excitation profiles for the states $|f\rangle$ (top panels) and $|e\rangle$ (bottom panels) in the $\mathrm{S}_\mathrm{a}$-NB5, $\mathrm{S}_\mathrm{a}$-PB5, $\mathrm{P}_\mathrm{a}$-NB5, and $\mathrm{P}_\mathrm{a}$-PB5b sequences.
All modulation parameters are presented in Table~\ref{tab5}.}\label{FIG.9}
\end{figure*}

The construction method of NB and PB sequences is readily generalized to achieve arbitrary population transfer between two lower states in the three-state system. In the following, we make use of the pulse number $N=5$ as examples to illustrate this issue.
To make a distinction,
we call the NB5 and PB5 sequences as the $\mathrm{S}_\mathrm{a}$-NB5 and $\mathrm{S}_\mathrm{a}$-PB5 sequences in the strength modulation,
while label them as the $\mathrm{P}_\mathrm{a}$-NB5 and $\mathrm{P}_\mathrm{a}$-PB5X sequences in the phase modulation.

\renewcommand\arraystretch{1.2}
\begin{table}[t]
\caption{Coefficients to be nullified for different types of sequences.\label{tab6}}
\setlength\tabcolsep{19.8pt}
 \centering
 \begin{tabular}{ll}
 \hline
  \hline
Sequence&Coefficients\\
  \hline
  $\mathrm{S}_\mathrm{a}$-NB5&$\tilde{x}_{5,4}~~~\tilde{x}_{5,6}~~~y_{5,2}~~~\tilde{y}_{5,2}$\\[-1ex]
$\mathrm{S}_\mathrm{a}$-PB5&$x_{5,2}~~~\tilde{x}_{5,4}~~~y_{5,2}~~~\tilde{y}_{5,2}$\\[-1ex]
$\mathrm{P}_\mathrm{a}$-NB5&$\tilde{x}_{5,4}~~~\tilde{x}_{5,6}~~~\tilde{x}_{5,8}~~~\tilde{x}_{5,10}~~~y_{5,2}~~~y_{5,4}~~~\tilde{y}_{5,2}~~~\tilde{y}_{5,4}$\\[-1ex]
$\mathrm{P}_\mathrm{a}$-PB5b&$x_{5,2}~~~x_{5,4}~~~\tilde{x}_{5,4}~~~~\!\tilde{x}_{5,6}~~~~y_{5,2}~~~y_{5,4}~~~\tilde{y}_{5,2}~~~\tilde{y}_{5,4}$\\
  \hline
  \hline
 \end{tabular}
\end{table}

First,
Eq.~(\ref{populationinversion}) outlines the primary condition for achieving arbitrary population transfer.
The design procedure is slightly distinct from those studied in Sec.~\ref{pulse design}.
Because part of modulation parameters are used to nullify the derivatives in the transition probability $P^{(N)}_e(\epsilon)$ according to Eq.~(\ref{leakeq}),
in the same pulse number,
there are few types to the PB sequences for arbitrary population transfer.
For instance,
in the $\mathrm{S}_\mathrm{a}$-PB5 sequence,
four coupling strength ratios are utilized to cancel out $y_{5,2}$, $\tilde{y}_{5,2}$, $x_{5,2}$, and $\tilde{x}_{5,4}$,
resulting in a single type of the PB sequence.
To make it more intuitive,
we present in Table~\ref{tab6} all coefficients to be nullified for establishing the $\mathrm{S}_\mathrm{a}$-NB5,
$\mathrm{S}_\mathrm{a}$-PB5,
$\mathrm{P}_\mathrm{a}$-NB5,
and $\mathrm{P}_\mathrm{a}$-PB5b sequences,
where their solutions are given by Table~\ref{tab5}.

\renewcommand\arraystretch{1.15}
\begin{table}[hp]
\centering
\caption{
Coupling strength ratios and phases for the $\mathrm{S}_\mathrm{a}$-$\mathrm{NB}5$, $\mathrm{S}_\mathrm{a}$-$\mathrm{PB}5$, $\mathrm{P}_\mathrm{a}$-$\mathrm{NB}5$, and $\mathrm{P}_\mathrm{a}$-PB5b sequences (in units of $\pi$).\label{tab5}}
\setlength\tabcolsep{4.5pt}
 \begin{tabular}{ccccccccccccccccccccc}
 \hline
 \hline
$\mathcal{P}$&0.9 &0.8&0.7&0.6&0.5&0.4&0.3&0.2&0.1\\
 \hline
$\mathrm{S}_\mathrm{a}$-NB5&\\[-1ex]
 $\theta_1$&0.1216&1.1226&1.1233&1.1236&1.8763&0.8765&1.1227&0.8788&0.1174\\[-1.5ex]
 $\theta_2$&1.7835&0.7679&0.7543&0.7410&0.2726&1.2876&0.6951&1.3270&1.6390\\[-1.5ex]
 $\theta_3$&1.1930&0.1478&0.1096&0.0738&0.9620&1.9996&1.9586&0.0918&0.8365\\[-1.5ex]
 $\theta_4$&1.9769&0.9344&0.8987&0.8656&0.1673&1.2015&0.7609&1.2839&1.6533\\[-1.5ex]
 $\theta_5$&0.6446&1.6080&1.5779&1.5502&1.4766&0.5039&1.4669&0.5665&0.3896\\
   \hline
$\mathrm{S}_\mathrm{a}$-PB5&\\[-1ex]
 $\theta_1$&1.2200&1.2324&1.7581&1.7499&0.2577&1.2650&0.2725&1.2808&0.9936\\[-1.5ex]
 $\theta_2$&1.6545&1.7206&1.2290&1.1850&0.8562&1.8972&0.9403&1.9897&0.9936\\[-1.5ex]
 $\theta_3$&0.9604&1.0559&1.8707&1.8065&0.2536&1.3130&0.3750&1.4447&1.0320\\[-1.5ex]
 $\theta_4$&0.1420&0.2114&0.7332&0.6835&1.3639&0.4115&1.4617&0.5189&1.0320\\[-1.5ex]
 $\theta_5$&0.4173&0.4674&0.4912&0.4532&1.5839&0.6217&1.6623&0.7093&0.0512\\
\hline
$\mathrm{P}_\mathrm{a}$-NB5&\\[-1ex]
       $\theta$&0.8424&1.8321&1.1578&0.8281&1.1242&1.0925&1.8988&1.8778&0.0285\\[-1.5ex]
      $\phi_2$&1.4751&1.2427&0.6861&1.0835&1.3505&0.7847&1.3257&1.3005&0.8716\\[-1.5ex]
 $\varphi_2$&0.1375&0.9586&1.0145&1.9725&0.8904&1.0196&0.1656&0.9449&1.1301\\[-1.5ex]
      $\phi_3$&1.3655&0.5775&1.1615&1.9709&0.5951&1.3670&1.8897&0.7344&1.3311\\[-1.5ex]
 $\varphi_3$&1.1045&1.7474&0.1907&0.9366&1.9091&0.2470&1.1773&1.8017&0.7655\\[-1.5ex]
      $\phi_4$&0.5345&1.0085&1.3687&0.5786&0.6205&0.7631&0.8998&0.5405&1.8695\\[-1.5ex]
 $\varphi_4$&0.2872&1.8213&1.9197&0.8735&0.1019&0.1208&0.5262&1.5857&0.1531\\[-1.5ex]
      $\phi_5$&0.6323&1.9165&0.1921&1.2124&1.5372&1.8727&0.6552&1.6259&0.6640\\[-1.5ex]
 $\varphi_5$&1.0487&1.0543&1.0879&1.9511&1.2865&0.9275&1.5082&0.4730&1.3318\\
\hline
    $\mathrm{P}_\mathrm{a}$-PB5b&\\[-1ex]
       $\theta$&0.1988&1.8236&0.8423&1.1410&1.8780&1.8910&0.0923&0.1194&0.9821\\[-1.5ex]
      $\phi_2$&0.7756&1.2414&0.6145&1.4103&0.2878&0.2254&1.3906&1.3457&0.4795\\[-1.5ex]
 $\varphi_2$&0.7756&1.2395&0.6145&1.4103&1.7025&0.2254&1.3906&0.5910&1.7838\\[-1.5ex]
      $\phi_3$&1.1354&0.8789&1.0493&0.9612&0.9553&0.9285&0.9585&0.6404&1.0012\\[-1.5ex]
 $\varphi_3$&0.0731&1.9549&1.7835&0.2469&0.8287&1.3337&0.2075&1.4067&1.2750\\[-1.5ex]
      $\phi_4$&1.5089&0.5249&1.2708&0.7553&1.6212&0.9666&0.7334&0.5750&1.8134\\[-1.5ex]
 $\varphi_4$&0.4466&1.5980&0.0050&0.0411&1.1949&1.3717&1.9824&1.3433&0.9507\\[-1.5ex]
      $\phi_5$&0.2770&1.7618&1.9905&0.0459&0.9992&1.5673&0.0218&1.5403&1.1453\\[-1.5ex]
 $\varphi_5$&1.6150&0.4147&1.1533&0.8822&0.0413&0.5503&0.7901&0.2147&0.1682\\
  \hline
  \hline
 \end{tabular}
\end{table}

In Figs.~\ref{FIG.9}(a)-(d), we plot different excitation profiles of the target state $|f\rangle$ for the $\mathrm{S}_\mathrm{a}$-NB5,
$\mathrm{S}_\mathrm{a}$-PB5,
$\mathrm{P}_\mathrm{a}$-NB5,
and $\mathrm{P}_\mathrm{a}$-PB5b sequences,
respectively,
while the corresponding population of the third state $|e\rangle$ are presented in Figs.~\ref{FIG.9}(e)-(h).
As expected, Figs.~\ref{FIG.9}(a)-(d) verify that the current design is successful in achieving arbitrary population transfer for different types of NB and PB sequences.
Moreover,
Figs.~\ref{FIG.9}(e)-(h) demonstrate the current sequences have remarkable ability for suppressing leakage,
because the population of the excited state is always restricted at an extremely low level for all $\epsilon$.

\section{Influence of imperfect pulses}\label{environment}

This section examines the effectiveness of the current sequences when the pulses are imperfect.
Imperfect pulses, such as waveform distortion and phase or detuning offset, are usually caused by defective experimental devices, imprecise operations, external noises, etc.
In the following,
we use the S-NB5 and P-NB5 sequences to investigate the impact of imperfect pulses.

\subsection{Waveform distortion}

\begin{figure}[t]
\centering
\scalebox{0.6}{\includegraphics{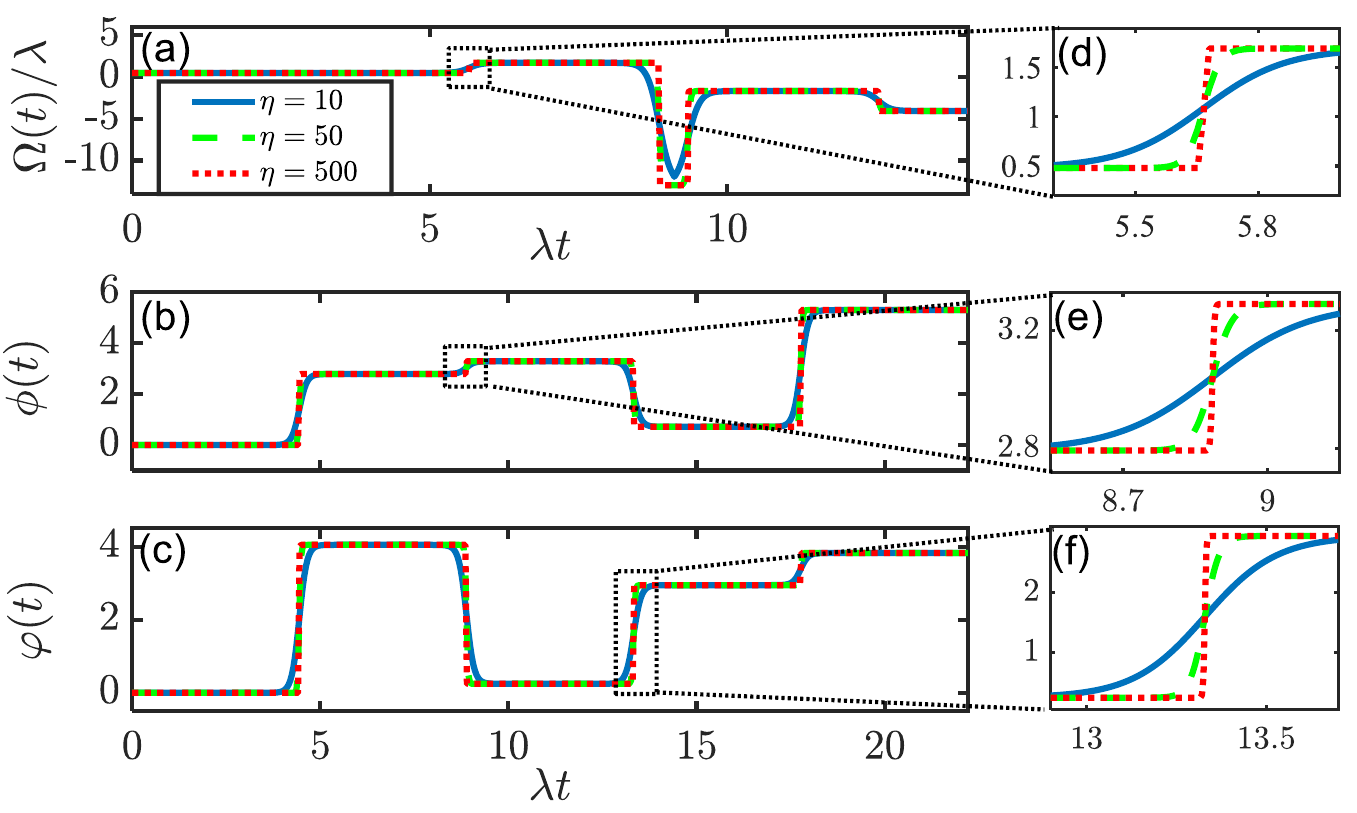}}
\caption{Different waveforms of the function $\alpha(t)$. Top: the coupling strength $\Omega(t)$. Middle: the phase $\phi(t)$. Bottom: the phase $\varphi(t)$.
}\label{FIG.12}
\end{figure}

In reality,
due to the inability of instantaneously switching between different values,
the unknown rising and falling edges inevitably appear in the square wave.
To visualize the influence of waveform distortion on the performance of the transition probability,
we utilize a time-dependent function to simulate the square wave, whose form is expressed as
\begin{eqnarray}
\alpha(t)=\frac{\alpha_n\!-\!\alpha_{n-1}}{1\!+\!\exp{\left[\eta(t\!-\!\sum^{n-1}_{l=1}t_{l})\right]}}, ~~\tau_{n-2}<t\leq\tau_{n-1},~~~~
\end{eqnarray}
with
\begin{eqnarray}
\tau_0=0,  ~~\tau_k=\sum_{l=1}^{k}t_l+\frac{t_{k+1}}{2}, ~~\tau_4=T, ~~k=1,2,3, \nonumber
\end{eqnarray}
where $\eta$ is a dimensionless parameter for controlling the magnitude of the rising and falling edges.
Here, $\alpha(t)$ denotes the expression of the coupling strength $\Omega(t)$ in the strength modulation,
while represents $\phi(t)$ and $\varphi(t)$ in the phase modulation.
We plot in Figs.~\ref{FIG.12}(a)-(c) the waveforms of $\alpha(t)$ with different $\eta$.
It is shown in Figs.~\ref{FIG.12}(d)-(f) that
the function $\alpha(t)$ gets closer to a perfect square wave as the parameter $\eta$ increases.
For example,
the function $\alpha(t)$ can be almost regarded as the square wave at $\eta=500$.
Note that the function $\alpha(t)$ becomes a smooth curve when $\eta$ is small meaning that there is severe waveform distortion in the square wave.

As shown in Figs.~\ref{FIG.13r}(a) and \ref{FIG.13r}(b),
the S-NB5 sequence possesses a well performance even in the presence of severe waveform distortion ($\eta\sim10$),
but the effectiveness of the P-NB5 sequence highly depends on the waveform.
Applying a near-perfect square waveform ($\eta\sim100$) can greatly protect the selectivity of the P-NB5 sequence from the pulse area error.
The reason is that the phase modulation relies heavily on high-precision phases so as to ensure the operational accuracy \cite{PhysRevA.103.033110}.

\subsection{Phase errors}

Previous studies \cite{PhysRevA.103.033110,Shi2022} have demonstrated that the effectiveness of the CP design may be hindered by phase errors.
It becomes particularly obvious in the phase modulation,
since the accuracy of phases directly impacts the waveform of sequences \cite{PhysRevA.100.023410}.
Therefore,
it is imperative to evaluate the validity of the current method when the phases are inaccurate.

\begin{figure}[t]
\centering
\scalebox{0.50}{\includegraphics{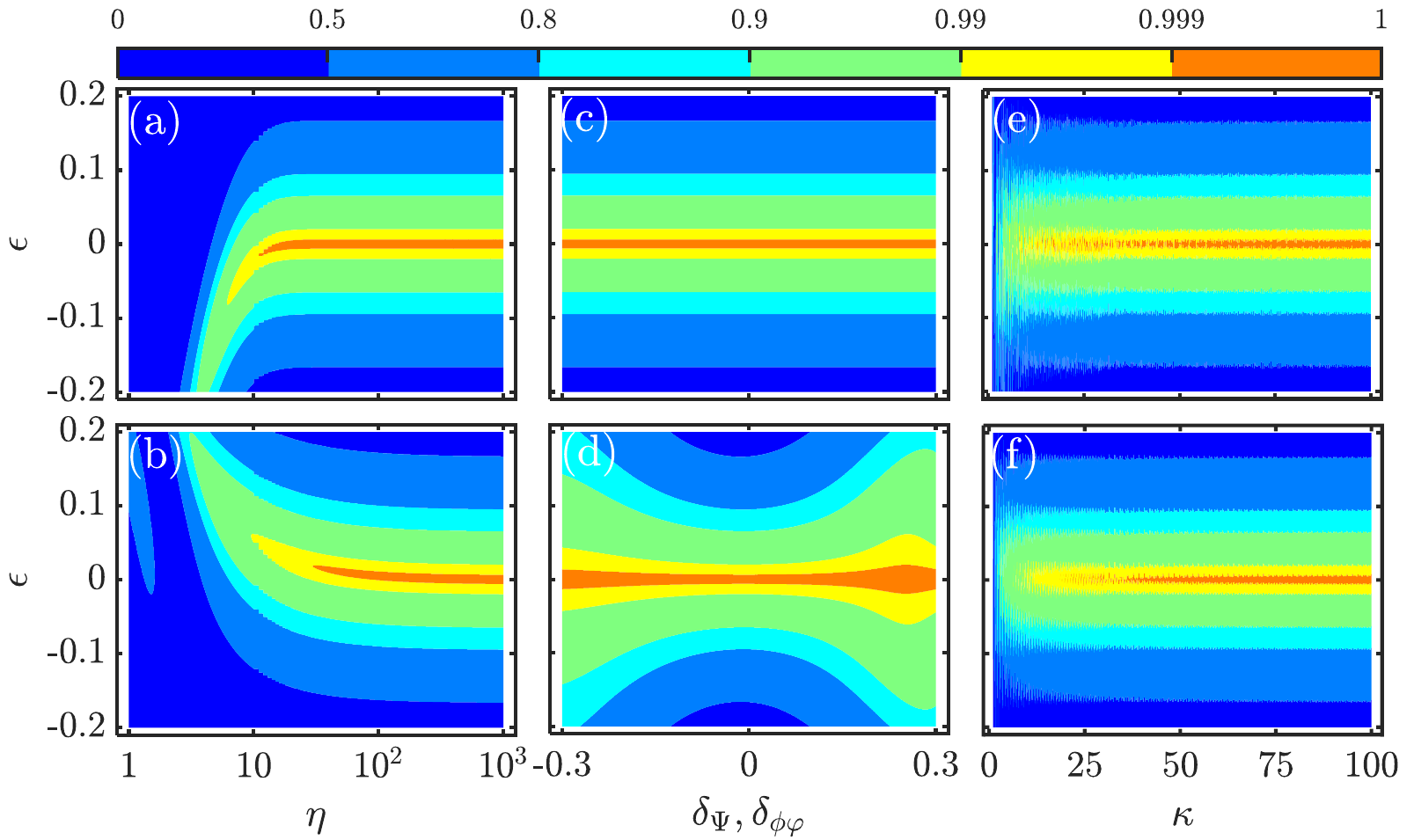}}
\caption{Excitation profiles in the presence of different types of errors.
Left column: $P^{(5)}_f(\epsilon, \eta)$ vs $\epsilon$ and $\eta$.
Middle column: $P^{(5)}_f(\epsilon,\delta_{\Psi/\phi\varphi})$ vs $\epsilon$ and $\delta_{\Psi/\phi\varphi}$.
Right column: $P^{(5)}_f(\epsilon,\omega_f)$ vs $\epsilon$ and $\kappa$.
The top and bottom rows denote the S-NB5 and P-NB5 sequences,
respectively.
Here, $\eta=1000$ in panels (c)-(f),
$\omega_g=1.2\omega_f$ in panels (e) and (f),
and other parameters can be found in Tables~\ref{tab1} and \ref{tab3}.}\label{FIG.13r}
\end{figure}

In the strength modulation,
the phase error is assumed to exhibit in the phase difference $\Psi_n$ of each pulse, i.e.,
\begin{eqnarray}
\Psi_n\rightarrow\Psi_n(1+\delta_{\Psi}),
\end{eqnarray}
while it appears in $\phi_n$ and $\varphi_n$ in the phase modulation,
\begin{eqnarray}
\phi_n \rightarrow \phi_n(1+\delta_{\phi}),~~~\varphi_n \rightarrow \varphi_n(1+\delta_{\varphi}).
\end{eqnarray}
Figures~\ref{FIG.13r}(c) and \ref{FIG.13r}(d) demonstrate the transition probability of the state $|f\rangle$ as a function of the pulse area error and the phase error for the S-NB5 and P-NB5 sequences,
respectively,
where we set $\delta_{\phi}=\delta_{\varphi}=\delta_{\phi\varphi}$ for simplicity.
A very striking feature in Fig.~\ref{FIG.13r}(c) is that the phase error does not impact the excitation profile in the strength modulation.
This is due to the fact that phase errors become global when the phase difference $\Psi_n$ is the same for each pulse.
As a result,
the S-NB sequence is inherently robust against phase errors.
In regards to the phase modulation,
the high excitation region experiences a slight change, as illustrated by the orange region in Fig.~\ref{FIG.13r}(d).
The reason is that the inaccurate phases do not satisfy Eqs.~(\ref{eqNB}) any more,
resulting in the low-order error coefficients not being fully eliminated.
Therefore,
when the system exhibits phase errors,
the strength modulation outperforms the phase modulation.

\subsection{Effectiveness beyond RWA}

In some quantum systems, such as superconducting circuits \cite{RevModPhys.85.623},
the ultra-strong coupling regime is possible to attain under the state-of-art technology \cite{Niemczyk2010,PhysRevA.100.043815}.
Namely, the coupling strength is comparable to the transition frequency.
In this situation, the counter-rotating terms ignored in RWA can have a notable effect on the dynamics.
Next,
we investigate the effectiveness of the pulse sequences beyond RWA.

To this end,
we retain the counter-rotating terms in Hamiltonian~(\ref{Hami}).
We use a dimensionless parameter $\kappa$ to denote the ratio of the frequency $\omega_{f}$ and the coupling strength $\lambda$,
which reads $\kappa=\omega_{f}/\lambda$.
Obviously,
a large $\kappa$ means that the transition frequency is far large than the coupling strength so that the RWA can be justified.
In Figs.~\ref{FIG.13r}(e) and \ref{FIG.13r}(f),
we present the excitation profiles for the S-NB5 and P-NB5 sequences as a function of the error $\epsilon$ and the parameter $\kappa$.
It can be observed that the excitation profiles for both sequences exhibit significant deformation at small $\kappa$ ($\kappa<5$), because the RWA cannot be satisfied,
leading to invalidness of the NB sequence.
As the value of $\kappa$ increases,
both sequences perform well.
And the phase modulation possesses better robustness to the influence caused by the counter-rotating terms.
To be specific,
the P-NB5 sequence becomes more capable of generating a narrow excitation profile at a moderate $\kappa$ ($\kappa \sim 25$),
while the S-NB5 sequence requires a higher $\kappa$ ($\kappa> 40$) to achieve the same effectiveness.
The reason is that,
the coupling strengths in different sub-pulses are diverse from each other and thus cannot always be much smaller than the transition frequency in the strength modulation.
In this situation,
the unfavorable influence of the counter-rotating terms becomes obvious and is hard to be ignored.

\subsection{Detuning errors}

Detuning errors generally result from a discrepancy between the transition frequency and the driving field frequency,
and make the pulses become off-resonance.
In the following,
we study the subject of $\delta_{g(f)}\neq 0$.

\begin{figure}[t]
\centering
\scalebox{0.6}{\includegraphics{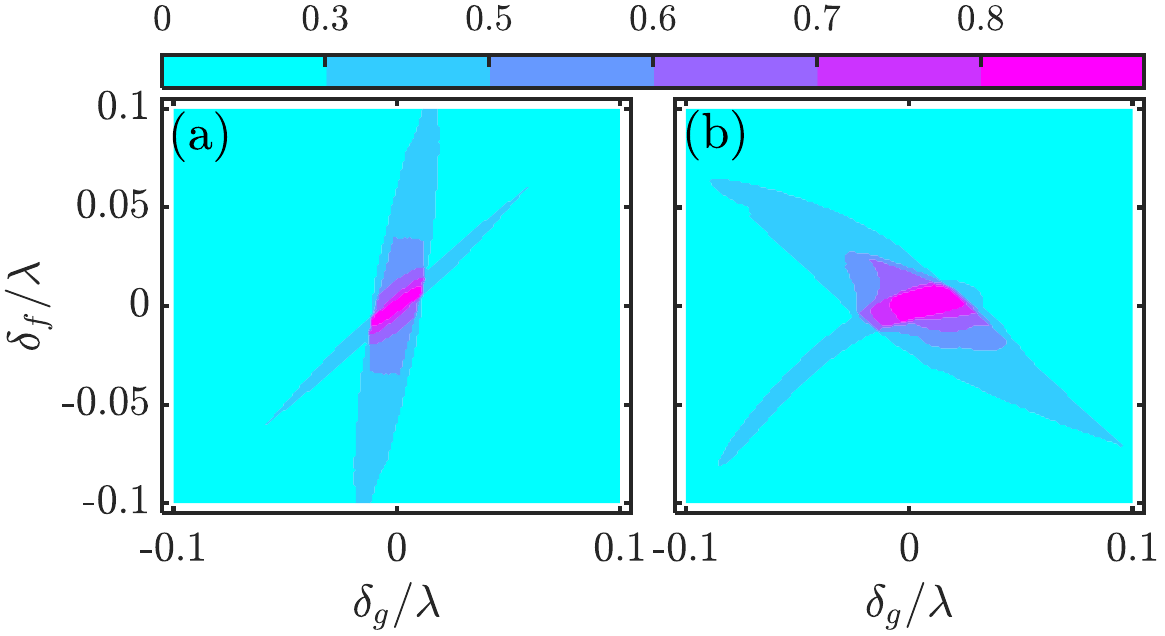}}
\caption{Low excitation width $W_l$ of the excitation profile for (a) S-NB5 and (b) P-NB5 sequences against the detunings $\delta_g$ and $\delta_f$. All parameters can be found in Tables~\ref{tab1} and \ref{tab3}.}\label{FIG.14}
\end{figure}

Figures~\ref{FIG.14}(a) and \ref{FIG.14}(b) demonstrate the low excitation width $W_l$ of the excitation profile for the S-NB5 and P-NB5 sequences at different $\delta_g$ and $\delta_f$.
The P-NB5 sequence exhibits better selectivity to excitation when a small detuning occurs,
while it is not satisfactory for the S-NB5 sequence, as indicated by the pink regions in Figs.~\ref{FIG.14}(a) and \ref{FIG.14}(b).
These results show that the strength modulation is sensitive to detuning errors in NB sequences.
In the strength modulation, $\Omega$ varies in the whole sequence.
Once the detuning is larger than the coupling strength,
population transfer cannot be successfully manipulated due to the large detuning.
As a consequence,
the selective excitation for the NB sequence would be dramatically affected.
On the other hand,
in the phase modulation,
the coupling strength consistently exceeds the detunings,
making the selective excitation less affected.

\section{Conclusion}\label{con}

In summary,
we propose a systematic method for constructing the NB and PB sequences that to implement efficient population inversion in the three-state system.
The modulation parameters of sequences are designed through either nullifying the coefficients of error terms or minimizing the cost function.
The former is typically preferred due to its ability to create a perfect excitation profile,
with flatter top or bottoms.
However,
the latter offers greater universality as it can tackle the situation in which the former fails,
and still successfully shape a nearly perfect excitation profile.
The results shows that both the strength and phase modulations can produce the NB sequences with identical excitation profiles,
and the PB sequences with the desired shape of excitation profiles.

Moreover,
the current method can be also extended to create the NB and PB sequences for arbitrary population transfer.
These sequences exhibit remarkable suppression of leakage to the excited state.
It is very important that,
the sequences are robust against waveform distortion by using the strength modulation,
while perform well in off-resonance conditions by the phase modulation.
Furthermore, the design method remains valid for both modulation techniques beyond RWA,
where the phase modulation performs better.
We hope this work can offer a promising candidate for precise quantum control (e.g., the high-precision addressing operation by laser fields) in different physical systems.

\begin{acknowledgments}
This work is supported by the Natural Science Foundation of Fujian Province under Grant No.~2021J01575,
the Natural Science Funds for Distinguished Young Scholar of Fujian Province under Grant No.~2020J06011,
and the Project from Fuzhou University under Grant No.~JG202001-2.
\end{acknowledgments}

\bibliographystyle{apsrev4-1}
\bibliography{reference2}

\begin{thebibliography}{80}%
\makeatletter
\providecommand \@ifxundefined [1]{%
 \@ifx{#1\undefined}
}%
\providecommand \@ifnum [1]{%
 \ifnum #1\expandafter \@firstoftwo
 \else \expandafter \@secondoftwo
 \fi
}%
\providecommand \@ifx [1]{%
 \ifx #1\expandafter \@firstoftwo
 \else \expandafter \@secondoftwo
 \fi
}%
\providecommand \natexlab [1]{#1}%
\providecommand \enquote  [1]{``#1''}%
\providecommand \bibnamefont  [1]{#1}%
\providecommand \bibfnamefont [1]{#1}%
\providecommand \citenamefont [1]{#1}%
\providecommand \href@noop [0]{\@secondoftwo}%
\providecommand \href [0]{\begingroup \@sanitize@url \@href}%
\providecommand \@href[1]{\@@startlink{#1}\@@href}%
\providecommand \@@href[1]{\endgroup#1\@@endlink}%
\providecommand \@sanitize@url [0]{\catcode `\\12\catcode `\$12\catcode
  `\&12\catcode `\#12\catcode `\^12\catcode `\_12\catcode `\%12\relax}%
\providecommand \@@startlink[1]{}%
\providecommand \@@endlink[0]{}%
\providecommand \url  [0]{\begingroup\@sanitize@url \@url }%
\providecommand \@url [1]{\endgroup\@href {#1}{\urlprefix }}%
\providecommand \urlprefix  [0]{URL }%
\providecommand \Eprint [0]{\href }%
\providecommand \doibase [0]{http://dx.doi.org/}%
\providecommand \selectlanguage [0]{\@gobble}%
\providecommand \bibinfo  [0]{\@secondoftwo}%
\providecommand \bibfield  [0]{\@secondoftwo}%
\providecommand \translation [1]{[#1]}%
\providecommand \BibitemOpen [0]{}%
\providecommand \bibitemStop [0]{}%
\providecommand \bibitemNoStop [0]{.\EOS\space}%
\providecommand \EOS [0]{\spacefactor3000\relax}%
\providecommand \BibitemShut  [1]{\csname bibitem#1\endcsname}%
\let\auto@bib@innerbib\@empty
\bibitem [{\citenamefont {Levitt}\ and\ \citenamefont
  {Freeman}(1979)}]{Levitt1979}%
  \BibitemOpen
  \bibfield  {author} {\bibinfo {author} {\bibfnamefont {M.~H.}\ \bibnamefont
  {Levitt}}\ and\ \bibinfo {author} {\bibfnamefont {R.}~\bibnamefont
  {Freeman}},\ }\bibinfo {title} {{NMR} population inversion using a composite
  pulse},\ \href {\doibase 10.1016/0022-2364(79)90265-8} {\bibfield  {journal}
  {\bibinfo  {journal} {J. Magn. Reson.}\ }\textbf {\bibinfo {volume} {33}},\
  \bibinfo {pages} {473} (\bibinfo {year} {1979})}\BibitemShut {NoStop}%
\bibitem [{\citenamefont {Wimperis}(1990)}]{WIMPERIS199046}%
  \BibitemOpen
  \bibfield  {author} {\bibinfo {author} {\bibfnamefont {S.}~\bibnamefont
  {Wimperis}},\ }\bibinfo {title} {Broadband and narrowband composite
  excitation sequences},\ \href {\doibase
  https://doi.org/10.1016/0022-2364(90)90210-Z} {\bibfield  {journal} {\bibinfo
   {journal} {J. Magn. Reson.}\ }\textbf {\bibinfo {volume} {86}},\ \bibinfo
  {pages} {46} (\bibinfo {year} {1990})}\BibitemShut {NoStop}%
\bibitem [{\citenamefont {Wimperis}(1989)}]{WIMPERIS1989509}%
  \BibitemOpen
  \bibfield  {author} {\bibinfo {author} {\bibfnamefont {S.}~\bibnamefont
  {Wimperis}},\ }\bibinfo {title} {Composite pulses with rectangular excitation
  and inversion profiles},\ \href {\doibase
  https://doi.org/10.1016/0022-2364(89)90346-6} {\bibfield  {journal} {\bibinfo
   {journal} {J. Magn. Reson.}\ }\textbf {\bibinfo {volume} {83}},\ \bibinfo
  {pages} {509} (\bibinfo {year} {1989})}\BibitemShut {NoStop}%
\bibitem [{\citenamefont {Tycko}\ \emph {et~al.}(1985)\citenamefont {Tycko},
  \citenamefont {Cho}, \citenamefont {Schneider},\ and\ \citenamefont
  {Pines}}]{TYCKO198590}%
  \BibitemOpen
  \bibfield  {author} {\bibinfo {author} {\bibfnamefont {R.}~\bibnamefont
  {Tycko}}, \bibinfo {author} {\bibfnamefont {H.}~\bibnamefont {Cho}}, \bibinfo
  {author} {\bibfnamefont {E.}~\bibnamefont {Schneider}}, \ and\ \bibinfo
  {author} {\bibfnamefont {A.}~\bibnamefont {Pines}},\ }\bibinfo {title}
  {Composite pulses without phase distortion},\ \href {\doibase
  https://doi.org/10.1016/0022-2364(85)90270-7} {\bibfield  {journal} {\bibinfo
   {journal} {J. Magn. Reson.}\ }\textbf {\bibinfo {volume} {61}},\ \bibinfo
  {pages} {90} (\bibinfo {year} {1985})}\BibitemShut {NoStop}%
\bibitem [{\citenamefont {Levitt}(1986)}]{Levitt1986}%
  \BibitemOpen
  \bibfield  {author} {\bibinfo {author} {\bibfnamefont {M.~H.}\ \bibnamefont
  {Levitt}},\ }\bibinfo {title} {Composite pulses},\ \href {\doibase
  10.1016/0079-6565(86)80005-x} {\bibfield  {journal} {\bibinfo  {journal}
  {Prog. NMR Spectrosc.}\ }\textbf {\bibinfo {volume} {18}},\ \bibinfo {pages}
  {61} (\bibinfo {year} {1986})}\BibitemShut {NoStop}%
\bibitem [{\citenamefont {Torosov}\ and\ \citenamefont
  {Vitanov}(2022)}]{PhysRevApplied.18.034062}%
  \BibitemOpen
  \bibfield  {author} {\bibinfo {author} {\bibfnamefont {B.~T.}\ \bibnamefont
  {Torosov}}\ and\ \bibinfo {author} {\bibfnamefont {N.~V.}\ \bibnamefont
  {Vitanov}},\ }\bibinfo {title} {Experimental Demonstration of Composite
  Pulses on IBM's Quantum Computer},\ \href {\doibase
  10.1103/PhysRevApplied.18.034062} {\bibfield  {journal} {\bibinfo  {journal}
  {Phys. Rev. Appl.}\ }\textbf {\bibinfo {volume} {18}},\ \bibinfo {pages}
  {034062} (\bibinfo {year} {2022})}\BibitemShut {NoStop}%
\bibitem [{\citenamefont {Ivanov}\ \emph {et~al.}(2022)\citenamefont {Ivanov},
  \citenamefont {Torosov},\ and\ \citenamefont
  {Vitanov}}]{PhysRevLett.129.240505}%
  \BibitemOpen
  \bibfield  {author} {\bibinfo {author} {\bibfnamefont {S.~S.}\ \bibnamefont
  {Ivanov}}, \bibinfo {author} {\bibfnamefont {B.~T.}\ \bibnamefont {Torosov}},
  \ and\ \bibinfo {author} {\bibfnamefont {N.~V.}\ \bibnamefont {Vitanov}},\
  }\bibinfo {title} {High-Fidelity Quantum Control by Polychromatic Pulse
  Trains},\ \href {\doibase 10.1103/PhysRevLett.129.240505} {\bibfield
  {journal} {\bibinfo  {journal} {Phys. Rev. Lett.}\ }\textbf {\bibinfo
  {volume} {129}},\ \bibinfo {pages} {240505} (\bibinfo {year}
  {2022})}\BibitemShut {NoStop}%
\bibitem [{\citenamefont {Gevorgyan}\ and\ \citenamefont
  {Vitanov}(2023)}]{PhysRevA.108.032614}%
  \BibitemOpen
  \bibfield  {author} {\bibinfo {author} {\bibfnamefont {H.~L.}\ \bibnamefont
  {Gevorgyan}}\ and\ \bibinfo {author} {\bibfnamefont {N.~V.}\ \bibnamefont
  {Vitanov}},\ }\bibinfo {title} {Deterministic generation of arbitrary
  ultrasmall excitation of quantum systems by composite pulse sequences},\
  \href {\doibase 10.1103/PhysRevA.108.032614} {\bibfield  {journal} {\bibinfo
  {journal} {Phys. Rev. A}\ }\textbf {\bibinfo {volume} {108}},\ \bibinfo
  {pages} {032614} (\bibinfo {year} {2023})}\BibitemShut {NoStop}%
\bibitem [{\citenamefont {Peters}\ \emph {et~al.}(2012)\citenamefont {Peters},
  \citenamefont {Ivanov}, \citenamefont {Englisch}, \citenamefont {Rangelov},
  \citenamefont {Vitanov},\ and\ \citenamefont {Halfmann}}]{Peters:12}%
  \BibitemOpen
  \bibfield  {author} {\bibinfo {author} {\bibfnamefont {T.}~\bibnamefont
  {Peters}}, \bibinfo {author} {\bibfnamefont {S.~S.}\ \bibnamefont {Ivanov}},
  \bibinfo {author} {\bibfnamefont {D.}~\bibnamefont {Englisch}}, \bibinfo
  {author} {\bibfnamefont {A.~A.}\ \bibnamefont {Rangelov}}, \bibinfo {author}
  {\bibfnamefont {N.~V.}\ \bibnamefont {Vitanov}}, \ and\ \bibinfo {author}
  {\bibfnamefont {T.}~\bibnamefont {Halfmann}},\ }\bibinfo {title} {Variable
  ultrabroadband and narrowband composite polarization retarders},\ \href
  {\doibase 10.1364/AO.51.007466} {\bibfield  {journal} {\bibinfo  {journal}
  {Appl. Opt.}\ }\textbf {\bibinfo {volume} {51}},\ \bibinfo {pages} {7466}
  (\bibinfo {year} {2012})}\BibitemShut {NoStop}%
\bibitem [{\citenamefont {Counsell}\ \emph {et~al.}(1985)\citenamefont
  {Counsell}, \citenamefont {Levitt},\ and\ \citenamefont
  {Ernst}}]{COUNSELL1985133}%
  \BibitemOpen
  \bibfield  {author} {\bibinfo {author} {\bibfnamefont {C.}~\bibnamefont
  {Counsell}}, \bibinfo {author} {\bibfnamefont {M.}~\bibnamefont {Levitt}}, \
  and\ \bibinfo {author} {\bibfnamefont {R.}~\bibnamefont {Ernst}},\ }\bibinfo
  {title} {Analytical theory of composite pulses},\ \href {\doibase
  https://doi.org/10.1016/0022-2364(85)90160-X} {\bibfield  {journal} {\bibinfo
   {journal} {J. Magn. Reson.}\ }\textbf {\bibinfo {volume} {63}},\ \bibinfo
  {pages} {133} (\bibinfo {year} {1985})}\BibitemShut {NoStop}%
\bibitem [{\citenamefont {Wimperis}(1991)}]{Wimperis1991}%
  \BibitemOpen
  \bibfield  {author} {\bibinfo {author} {\bibfnamefont {S.}~\bibnamefont
  {Wimperis}},\ }\bibinfo {title} {Iterative schemes for phase-distortionless
  composite 180{\textdegree} pulses},\ \href {\doibase
  10.1016/0022-2364(91)90043-s} {\bibfield  {journal} {\bibinfo  {journal} {J.
  Magn. Reson.}\ }\textbf {\bibinfo {volume} {93}},\ \bibinfo {pages} {199}
  (\bibinfo {year} {1991})}\BibitemShut {NoStop}%
\bibitem [{\citenamefont {Jones}(2011)}]{JONES201191}%
  \BibitemOpen
  \bibfield  {author} {\bibinfo {author} {\bibfnamefont {J.~A.}\ \bibnamefont
  {Jones}},\ }\bibinfo {title} {Quantum computing with {NMR}},\ \href {\doibase
  https://doi.org/10.1016/j.pnmrs.2010.11.001} {\bibfield  {journal} {\bibinfo
  {journal} {Prog. NMR Spectrosc.}\ }\textbf {\bibinfo {volume} {59}},\
  \bibinfo {pages} {91} (\bibinfo {year} {2011})}\BibitemShut {NoStop}%
\bibitem [{\citenamefont {Odedra}\ and\ \citenamefont
  {Wimperis}(2012)}]{ODEDRA201268}%
  \BibitemOpen
  \bibfield  {author} {\bibinfo {author} {\bibfnamefont {S.}~\bibnamefont
  {Odedra}}\ and\ \bibinfo {author} {\bibfnamefont {S.}~\bibnamefont
  {Wimperis}},\ }\bibinfo {title} {Use of composite refocusing pulses to form
  spin echoes},\ \href {\doibase https://doi.org/10.1016/j.jmr.2011.10.006}
  {\bibfield  {journal} {\bibinfo  {journal} {J. Magn. Reson.}\ }\textbf
  {\bibinfo {volume} {214}},\ \bibinfo {pages} {68} (\bibinfo {year}
  {2012})}\BibitemShut {NoStop}%
\bibitem [{\citenamefont {Odedra}\ \emph {et~al.}(2012)\citenamefont {Odedra},
  \citenamefont {Thrippleton},\ and\ \citenamefont {Wimperis}}]{ODEDRA201281}%
  \BibitemOpen
  \bibfield  {author} {\bibinfo {author} {\bibfnamefont {S.}~\bibnamefont
  {Odedra}}, \bibinfo {author} {\bibfnamefont {M.~J.}\ \bibnamefont
  {Thrippleton}}, \ and\ \bibinfo {author} {\bibfnamefont {S.}~\bibnamefont
  {Wimperis}},\ }\bibinfo {title} {Dual-compensated antisymmetric composite
  refocusing pulses for NMR},\ \href {\doibase
  https://doi.org/10.1016/j.jmr.2012.10.003} {\bibfield  {journal} {\bibinfo
  {journal} {J. Magn. Reson.}\ }\textbf {\bibinfo {volume} {225}},\ \bibinfo
  {pages} {81} (\bibinfo {year} {2012})}\BibitemShut {NoStop}%
\bibitem [{\citenamefont {Brown}\ \emph {et~al.}(2004)\citenamefont {Brown},
  \citenamefont {Harrow},\ and\ \citenamefont {Chuang}}]{PhysRevA.70.052318}%
  \BibitemOpen
  \bibfield  {author} {\bibinfo {author} {\bibfnamefont {K.~R.}\ \bibnamefont
  {Brown}}, \bibinfo {author} {\bibfnamefont {A.~W.}\ \bibnamefont {Harrow}}, \
  and\ \bibinfo {author} {\bibfnamefont {I.~L.}\ \bibnamefont {Chuang}},\
  }\bibinfo {title} {Arbitrarily accurate composite pulse sequences},\ \href
  {\doibase 10.1103/PhysRevA.70.052318} {\bibfield  {journal} {\bibinfo
  {journal} {Phys. Rev. A}\ }\textbf {\bibinfo {volume} {70}},\ \bibinfo
  {pages} {052318} (\bibinfo {year} {2004})}\BibitemShut {NoStop}%
\bibitem [{\citenamefont {Cummins}\ and\ \citenamefont
  {Jones}(2000)}]{Cummins_2000}%
  \BibitemOpen
  \bibfield  {author} {\bibinfo {author} {\bibfnamefont {H.~K.}\ \bibnamefont
  {Cummins}}\ and\ \bibinfo {author} {\bibfnamefont {J.~A.}\ \bibnamefont
  {Jones}},\ }\bibinfo {title} {Use of composite rotations to correct
  systematic errors in {NMR} quantum computation},\ \href {\doibase
  10.1088/1367-2630/2/1/006} {\bibfield  {journal} {\bibinfo  {journal} {New J.
  Phys.}\ }\textbf {\bibinfo {volume} {2}},\ \bibinfo {pages} {006} (\bibinfo
  {year} {2000})}\BibitemShut {NoStop}%
\bibitem [{\citenamefont {Said}\ and\ \citenamefont
  {Twamley}(2009)}]{PhysRevA.80.032303}%
  \BibitemOpen
  \bibfield  {author} {\bibinfo {author} {\bibfnamefont {R.~S.}\ \bibnamefont
  {Said}}\ and\ \bibinfo {author} {\bibfnamefont {J.}~\bibnamefont {Twamley}},\
  }\bibinfo {title} {Robust control of entanglement in a nitrogen-vacancy
  center coupled to a $^{13}\text{C}$ nuclear spin in diamond},\ \href
  {\doibase 10.1103/PhysRevA.80.032303} {\bibfield  {journal} {\bibinfo
  {journal} {Phys. Rev. A}\ }\textbf {\bibinfo {volume} {80}},\ \bibinfo
  {pages} {032303} (\bibinfo {year} {2009})}\BibitemShut {NoStop}%
\bibitem [{\citenamefont {Kabytayev}\ \emph {et~al.}(2014)\citenamefont
  {Kabytayev}, \citenamefont {Green}, \citenamefont {Khodjasteh}, \citenamefont
  {Biercuk}, \citenamefont {Viola},\ and\ \citenamefont
  {Brown}}]{PhysRevA.90.012316}%
  \BibitemOpen
  \bibfield  {author} {\bibinfo {author} {\bibfnamefont {C.}~\bibnamefont
  {Kabytayev}}, \bibinfo {author} {\bibfnamefont {T.~J.}\ \bibnamefont
  {Green}}, \bibinfo {author} {\bibfnamefont {K.}~\bibnamefont {Khodjasteh}},
  \bibinfo {author} {\bibfnamefont {M.~J.}\ \bibnamefont {Biercuk}}, \bibinfo
  {author} {\bibfnamefont {L.}~\bibnamefont {Viola}}, \ and\ \bibinfo {author}
  {\bibfnamefont {K.~R.}\ \bibnamefont {Brown}},\ }\bibinfo {title} {Robustness
  of composite pulses to time-dependent control noise},\ \href {\doibase
  10.1103/PhysRevA.90.012316} {\bibfield  {journal} {\bibinfo  {journal} {Phys.
  Rev. A}\ }\textbf {\bibinfo {volume} {90}},\ \bibinfo {pages} {012316}
  (\bibinfo {year} {2014})}\BibitemShut {NoStop}%
\bibitem [{\citenamefont {Dunning}\ \emph {et~al.}(2014)\citenamefont
  {Dunning}, \citenamefont {Gregory}, \citenamefont {Bateman}, \citenamefont
  {Cooper}, \citenamefont {Himsworth}, \citenamefont {Jones},\ and\
  \citenamefont {Freegarde}}]{PhysRevA.90.033608}%
  \BibitemOpen
  \bibfield  {author} {\bibinfo {author} {\bibfnamefont {A.}~\bibnamefont
  {Dunning}}, \bibinfo {author} {\bibfnamefont {R.}~\bibnamefont {Gregory}},
  \bibinfo {author} {\bibfnamefont {J.}~\bibnamefont {Bateman}}, \bibinfo
  {author} {\bibfnamefont {N.}~\bibnamefont {Cooper}}, \bibinfo {author}
  {\bibfnamefont {M.}~\bibnamefont {Himsworth}}, \bibinfo {author}
  {\bibfnamefont {J.~A.}\ \bibnamefont {Jones}}, \ and\ \bibinfo {author}
  {\bibfnamefont {T.}~\bibnamefont {Freegarde}},\ }\bibinfo {title} {Composite
  pulses for interferometry in a thermal cold atom cloud},\ \href {\doibase
  10.1103/PhysRevA.90.033608} {\bibfield  {journal} {\bibinfo  {journal} {Phys.
  Rev. A}\ }\textbf {\bibinfo {volume} {90}},\ \bibinfo {pages} {033608}
  (\bibinfo {year} {2014})}\BibitemShut {NoStop}%
\bibitem [{\citenamefont {Demeter}(2016)}]{PhysRevA.93.023830}%
  \BibitemOpen
  \bibfield  {author} {\bibinfo {author} {\bibfnamefont {G.}~\bibnamefont
  {Demeter}},\ }\bibinfo {title} {Composite pulses for high-fidelity population
  inversion in optically dense, inhomogeneously broadened atomic ensembles},\
  \href {\doibase 10.1103/PhysRevA.93.023830} {\bibfield  {journal} {\bibinfo
  {journal} {Phys. Rev. A}\ }\textbf {\bibinfo {volume} {93}},\ \bibinfo
  {pages} {023830} (\bibinfo {year} {2016})}\BibitemShut {NoStop}%
\bibitem [{\citenamefont {Torosov}\ and\ \citenamefont
  {Vitanov}(2013)}]{PhysRevA.87.043418}%
  \BibitemOpen
  \bibfield  {author} {\bibinfo {author} {\bibfnamefont {B.~T.}\ \bibnamefont
  {Torosov}}\ and\ \bibinfo {author} {\bibfnamefont {N.~V.}\ \bibnamefont
  {Vitanov}},\ }\bibinfo {title} {Composite stimulated Raman adiabatic
  passage},\ \href {\doibase 10.1103/PhysRevA.87.043418} {\bibfield  {journal}
  {\bibinfo  {journal} {Phys. Rev. A}\ }\textbf {\bibinfo {volume} {87}},\
  \bibinfo {pages} {043418} (\bibinfo {year} {2013})}\BibitemShut {NoStop}%
\bibitem [{\citenamefont {Dou}\ \emph {et~al.}(2016)\citenamefont {Dou},
  \citenamefont {Cao}, \citenamefont {Liu},\ and\ \citenamefont
  {Fu}}]{PhysRevA.93.043419}%
  \BibitemOpen
  \bibfield  {author} {\bibinfo {author} {\bibfnamefont {F.-Q.}\ \bibnamefont
  {Dou}}, \bibinfo {author} {\bibfnamefont {H.}~\bibnamefont {Cao}}, \bibinfo
  {author} {\bibfnamefont {J.}~\bibnamefont {Liu}}, \ and\ \bibinfo {author}
  {\bibfnamefont {L.-B.}\ \bibnamefont {Fu}},\ }\bibinfo {title} {High-fidelity
  composite adiabatic passage in nonlinear two-level systems},\ \href {\doibase
  10.1103/PhysRevA.93.043419} {\bibfield  {journal} {\bibinfo  {journal} {Phys.
  Rev. A}\ }\textbf {\bibinfo {volume} {93}},\ \bibinfo {pages} {043419}
  (\bibinfo {year} {2016})}\BibitemShut {NoStop}%
\bibitem [{\citenamefont {Al-Mahmoud}\ \emph {et~al.}(2020)\citenamefont
  {Al-Mahmoud}, \citenamefont {Coda}, \citenamefont {Rangelov},\ and\
  \citenamefont {Montemezzani}}]{PhysRevApplied.13.014048}%
  \BibitemOpen
  \bibfield  {author} {\bibinfo {author} {\bibfnamefont {M.}~\bibnamefont
  {Al-Mahmoud}}, \bibinfo {author} {\bibfnamefont {V.}~\bibnamefont {Coda}},
  \bibinfo {author} {\bibfnamefont {A.}~\bibnamefont {Rangelov}}, \ and\
  \bibinfo {author} {\bibfnamefont {G.}~\bibnamefont {Montemezzani}},\
  }\bibinfo {title} {Broadband polarization rotator with tunable rotation angle
  composed of three wave plates},\ \href {\doibase
  10.1103/PhysRevApplied.13.014048} {\bibfield  {journal} {\bibinfo  {journal}
  {Phys. Rev. Appl.}\ }\textbf {\bibinfo {volume} {13}},\ \bibinfo {pages}
  {014048} (\bibinfo {year} {2020})}\BibitemShut {NoStop}%
\bibitem [{\citenamefont {Zanon-Willette}\ \emph {et~al.}(2022)\citenamefont
  {Zanon-Willette}, \citenamefont {Wilkowski}, \citenamefont {Lefevre},
  \citenamefont {Taichenachev},\ and\ \citenamefont
  {Yudin}}]{PhysRevResearch.4.023222}%
  \BibitemOpen
  \bibfield  {author} {\bibinfo {author} {\bibfnamefont {T.}~\bibnamefont
  {Zanon-Willette}}, \bibinfo {author} {\bibfnamefont {D.}~\bibnamefont
  {Wilkowski}}, \bibinfo {author} {\bibfnamefont {R.}~\bibnamefont {Lefevre}},
  \bibinfo {author} {\bibfnamefont {A.~V.}\ \bibnamefont {Taichenachev}}, \
  and\ \bibinfo {author} {\bibfnamefont {V.~I.}\ \bibnamefont {Yudin}},\
  }\bibinfo {title} {Generalized hyper-Ramsey-Bord\'e matter-wave
  interferometry: Quantum engineering of robust atomic sensors with composite
  pulses},\ \href {\doibase 10.1103/PhysRevResearch.4.023222} {\bibfield
  {journal} {\bibinfo  {journal} {Phys. Rev. Research}\ }\textbf {\bibinfo
  {volume} {4}},\ \bibinfo {pages} {023222} (\bibinfo {year}
  {2022})}\BibitemShut {NoStop}%
\bibitem [{\citenamefont {Bodenstedt}\ \emph {et~al.}(2022)\citenamefont
  {Bodenstedt}, \citenamefont {Mitchell},\ and\ \citenamefont
  {Tayler}}]{PhysRevA.106.033102}%
  \BibitemOpen
  \bibfield  {author} {\bibinfo {author} {\bibfnamefont {S.}~\bibnamefont
  {Bodenstedt}}, \bibinfo {author} {\bibfnamefont {M.~W.}\ \bibnamefont
  {Mitchell}}, \ and\ \bibinfo {author} {\bibfnamefont {M.~C.~D.}\ \bibnamefont
  {Tayler}},\ }\bibinfo {title} {Meridional composite pulses for low-field
  magnetic resonance},\ \href {\doibase 10.1103/PhysRevA.106.033102} {\bibfield
   {journal} {\bibinfo  {journal} {Phys. Rev. A}\ }\textbf {\bibinfo {volume}
  {106}},\ \bibinfo {pages} {033102} (\bibinfo {year} {2022})}\BibitemShut
  {NoStop}%
\bibitem [{\citenamefont {Zhou}\ \emph {et~al.}(2021)\citenamefont {Zhou},
  \citenamefont {Li}, \citenamefont {Pan}, \citenamefont {Zhang}, \citenamefont
  {Chen},\ and\ \citenamefont {Xue}}]{PhysRevA.103.032609}%
  \BibitemOpen
  \bibfield  {author} {\bibinfo {author} {\bibfnamefont {J.}~\bibnamefont
  {Zhou}}, \bibinfo {author} {\bibfnamefont {S.}~\bibnamefont {Li}}, \bibinfo
  {author} {\bibfnamefont {G.-Z.}\ \bibnamefont {Pan}}, \bibinfo {author}
  {\bibfnamefont {G.}~\bibnamefont {Zhang}}, \bibinfo {author} {\bibfnamefont
  {T.}~\bibnamefont {Chen}}, \ and\ \bibinfo {author} {\bibfnamefont {Z.-Y.}\
  \bibnamefont {Xue}},\ }\bibinfo {title} {Nonadiabatic geometric quantum gates
  that are insensitive to qubit-frequency drifts},\ \href {\doibase
  10.1103/PhysRevA.103.032609} {\bibfield  {journal} {\bibinfo  {journal}
  {Phys. Rev. A}\ }\textbf {\bibinfo {volume} {103}},\ \bibinfo {pages}
  {032609} (\bibinfo {year} {2021})}\BibitemShut {NoStop}%
\bibitem [{\citenamefont {Chen}\ and\ \citenamefont
  {Xue}(2018)}]{PhysRevApplied.10.054051}%
  \BibitemOpen
  \bibfield  {author} {\bibinfo {author} {\bibfnamefont {T.}~\bibnamefont
  {Chen}}\ and\ \bibinfo {author} {\bibfnamefont {Z.-Y.}\ \bibnamefont {Xue}},\
  }\bibinfo {title} {Nonadiabatic geometric quantum computation with
  parametrically tunable coupling},\ \href {\doibase
  10.1103/PhysRevApplied.10.054051} {\bibfield  {journal} {\bibinfo  {journal}
  {Phys. Rev. Appl.}\ }\textbf {\bibinfo {volume} {10}},\ \bibinfo {pages}
  {054051} (\bibinfo {year} {2018})}\BibitemShut {NoStop}%
\bibitem [{\citenamefont {Dridi}\ \emph {et~al.}(2020)\citenamefont {Dridi},
  \citenamefont {Mejatty}, \citenamefont {Glaser},\ and\ \citenamefont
  {Sugny}}]{PhysRevA.101.012321}%
  \BibitemOpen
  \bibfield  {author} {\bibinfo {author} {\bibfnamefont {G.}~\bibnamefont
  {Dridi}}, \bibinfo {author} {\bibfnamefont {M.}~\bibnamefont {Mejatty}},
  \bibinfo {author} {\bibfnamefont {S.~J.}\ \bibnamefont {Glaser}}, \ and\
  \bibinfo {author} {\bibfnamefont {D.}~\bibnamefont {Sugny}},\ }\bibinfo
  {title} {Robust control of a NOT gate by composite pulses},\ \href {\doibase
  10.1103/PhysRevA.101.012321} {\bibfield  {journal} {\bibinfo  {journal}
  {Phys. Rev. A}\ }\textbf {\bibinfo {volume} {101}},\ \bibinfo {pages}
  {012321} (\bibinfo {year} {2020})}\BibitemShut {NoStop}%
\bibitem [{\citenamefont {Gevorgyan}\ and\ \citenamefont
  {Vitanov}(2021)}]{PhysRevA.104.012609}%
  \BibitemOpen
  \bibfield  {author} {\bibinfo {author} {\bibfnamefont {H.~L.}\ \bibnamefont
  {Gevorgyan}}\ and\ \bibinfo {author} {\bibfnamefont {N.~V.}\ \bibnamefont
  {Vitanov}},\ }\bibinfo {title} {Ultrahigh-fidelity composite rotational
  quantum gates},\ \href {\doibase 10.1103/PhysRevA.104.012609} {\bibfield
  {journal} {\bibinfo  {journal} {Phys. Rev. A}\ }\textbf {\bibinfo {volume}
  {104}},\ \bibinfo {pages} {012609} (\bibinfo {year} {2021})}\BibitemShut
  {NoStop}%
\bibitem [{\citenamefont {Ota}\ and\ \citenamefont
  {Kondo}(2009)}]{PhysRevA.80.024302}%
  \BibitemOpen
  \bibfield  {author} {\bibinfo {author} {\bibfnamefont {Y.}~\bibnamefont
  {Ota}}\ and\ \bibinfo {author} {\bibfnamefont {Y.}~\bibnamefont {Kondo}},\
  }\bibinfo {title} {Composite pulses in NMR as nonadiabatic geometric quantum
  gates},\ \href {\doibase 10.1103/PhysRevA.80.024302} {\bibfield  {journal}
  {\bibinfo  {journal} {Phys. Rev. A}\ }\textbf {\bibinfo {volume} {80}},\
  \bibinfo {pages} {024302} (\bibinfo {year} {2009})}\BibitemShut {NoStop}%
\bibitem [{\citenamefont {Shi}\ \emph {et~al.}(2022)\citenamefont {Shi},
  \citenamefont {Zhang}, \citenamefont {Ran}, \citenamefont {Xia},
  \citenamefont {Ianconescu}, \citenamefont {Friedman}, \citenamefont {Yi},\
  and\ \citenamefont {Zheng}}]{Shi2022}%
  \BibitemOpen
  \bibfield  {author} {\bibinfo {author} {\bibfnamefont {Z.-C.}\ \bibnamefont
  {Shi}}, \bibinfo {author} {\bibfnamefont {C.}~\bibnamefont {Zhang}}, \bibinfo
  {author} {\bibfnamefont {D.}~\bibnamefont {Ran}}, \bibinfo {author}
  {\bibfnamefont {Y.}~\bibnamefont {Xia}}, \bibinfo {author} {\bibfnamefont
  {R.}~\bibnamefont {Ianconescu}}, \bibinfo {author} {\bibfnamefont
  {A.}~\bibnamefont {Friedman}}, \bibinfo {author} {\bibfnamefont {X.~X.}\
  \bibnamefont {Yi}}, \ and\ \bibinfo {author} {\bibfnamefont {S.-B.}\
  \bibnamefont {Zheng}},\ }\bibinfo {title} {Composite pulses for high fidelity
  population transfer in three-level systems},\ \href {\doibase
  10.1088/1367-2630/ac48e7} {\bibfield  {journal} {\bibinfo  {journal} {New J.
  Phys.}\ }\textbf {\bibinfo {volume} {24}},\ \bibinfo {pages} {023014}
  (\bibinfo {year} {2022})}\BibitemShut {NoStop}%
\bibitem [{\citenamefont {Xu}\ \emph {et~al.}(2022)\citenamefont {Xu},
  \citenamefont {Song}, \citenamefont {Wang},\ and\ \citenamefont {Ye}}]{Xu22}%
  \BibitemOpen
  \bibfield  {author} {\bibinfo {author} {\bibfnamefont {H.}~\bibnamefont
  {Xu}}, \bibinfo {author} {\bibfnamefont {X.-K.}\ \bibnamefont {Song}},
  \bibinfo {author} {\bibfnamefont {D.}~\bibnamefont {Wang}}, \ and\ \bibinfo
  {author} {\bibfnamefont {L.}~\bibnamefont {Ye}},\ }\bibinfo {title} {Robust
  coherent control in three-level quantum systems using composite pulses},\
  \href {\doibase 10.1364/OE.449426} {\bibfield  {journal} {\bibinfo  {journal}
  {Opt. Express}\ }\textbf {\bibinfo {volume} {30}},\ \bibinfo {pages} {3125}
  (\bibinfo {year} {2022})}\BibitemShut {NoStop}%
\bibitem [{\citenamefont {Vandersypen}\ and\ \citenamefont
  {Chuang}(2005)}]{RevModPhys.76.1037}%
  \BibitemOpen
  \bibfield  {author} {\bibinfo {author} {\bibfnamefont {L.~M.~K.}\
  \bibnamefont {Vandersypen}}\ and\ \bibinfo {author} {\bibfnamefont {I.~L.}\
  \bibnamefont {Chuang}},\ }\bibinfo {title} {NMR techniques for quantum
  control and computation},\ \href {\doibase 10.1103/RevModPhys.76.1037}
  {\bibfield  {journal} {\bibinfo  {journal} {Rev. Mod. Phys.}\ }\textbf
  {\bibinfo {volume} {76}},\ \bibinfo {pages} {1037} (\bibinfo {year}
  {2005})}\BibitemShut {NoStop}%
\bibitem [{\citenamefont {Jones}(2013)}]{PhysRevA.87.052317}%
  \BibitemOpen
  \bibfield  {author} {\bibinfo {author} {\bibfnamefont {J.~A.}\ \bibnamefont
  {Jones}},\ }\bibinfo {title} {Designing short robust {NOT} gates for quantum
  computation},\ \href {\doibase 10.1103/PhysRevA.87.052317} {\bibfield
  {journal} {\bibinfo  {journal} {Phys. Rev. A}\ }\textbf {\bibinfo {volume}
  {87}},\ \bibinfo {pages} {052317} (\bibinfo {year} {2013})}\BibitemShut
  {NoStop}%
\bibitem [{\citenamefont {Torosov}\ \emph {et~al.}(2011)\citenamefont
  {Torosov}, \citenamefont {Gu\'erin},\ and\ \citenamefont
  {Vitanov}}]{PhysRevLett.106.233001}%
  \BibitemOpen
  \bibfield  {author} {\bibinfo {author} {\bibfnamefont {B.~T.}\ \bibnamefont
  {Torosov}}, \bibinfo {author} {\bibfnamefont {S.}~\bibnamefont {Gu\'erin}}, \
  and\ \bibinfo {author} {\bibfnamefont {N.~V.}\ \bibnamefont {Vitanov}},\
  }\bibinfo {title} {High-fidelity adiabatic passage by composite sequences of
  chirped pulses},\ \href {\doibase 10.1103/PhysRevLett.106.233001} {\bibfield
  {journal} {\bibinfo  {journal} {Phys. Rev. Lett.}\ }\textbf {\bibinfo
  {volume} {106}},\ \bibinfo {pages} {233001} (\bibinfo {year}
  {2011})}\BibitemShut {NoStop}%
\bibitem [{\citenamefont {Genov}\ \emph {et~al.}(2014)\citenamefont {Genov},
  \citenamefont {Schraft}, \citenamefont {Halfmann},\ and\ \citenamefont
  {Vitanov}}]{PhysRevLett.113.043001}%
  \BibitemOpen
  \bibfield  {author} {\bibinfo {author} {\bibfnamefont {G.~T.}\ \bibnamefont
  {Genov}}, \bibinfo {author} {\bibfnamefont {D.}~\bibnamefont {Schraft}},
  \bibinfo {author} {\bibfnamefont {T.}~\bibnamefont {Halfmann}}, \ and\
  \bibinfo {author} {\bibfnamefont {N.~V.}\ \bibnamefont {Vitanov}},\ }\bibinfo
  {title} {Correction of arbitrary field errors in population inversion of
  quantum systems by universal composite pulses},\ \href {\doibase
  10.1103/PhysRevLett.113.043001} {\bibfield  {journal} {\bibinfo  {journal}
  {Phys. Rev. Lett.}\ }\textbf {\bibinfo {volume} {113}},\ \bibinfo {pages}
  {043001} (\bibinfo {year} {2014})}\BibitemShut {NoStop}%
\bibitem [{\citenamefont {Low}\ \emph {et~al.}(2014)\citenamefont {Low},
  \citenamefont {Yoder},\ and\ \citenamefont {Chuang}}]{PhysRevA.89.022341}%
  \BibitemOpen
  \bibfield  {author} {\bibinfo {author} {\bibfnamefont {G.~H.}\ \bibnamefont
  {Low}}, \bibinfo {author} {\bibfnamefont {T.~J.}\ \bibnamefont {Yoder}}, \
  and\ \bibinfo {author} {\bibfnamefont {I.~L.}\ \bibnamefont {Chuang}},\
  }\bibinfo {title} {Optimal arbitrarily accurate composite pulse sequences},\
  \href {\doibase 10.1103/PhysRevA.89.022341} {\bibfield  {journal} {\bibinfo
  {journal} {Phys. Rev. A}\ }\textbf {\bibinfo {volume} {89}},\ \bibinfo
  {pages} {022341} (\bibinfo {year} {2014})}\BibitemShut {NoStop}%
\bibitem [{\citenamefont {Jones}(2003)}]{PhysRevA.67.012317}%
  \BibitemOpen
  \bibfield  {author} {\bibinfo {author} {\bibfnamefont {J.~A.}\ \bibnamefont
  {Jones}},\ }\bibinfo {title} {Robust Ising gates for practical quantum
  computation},\ \href {\doibase 10.1103/PhysRevA.67.012317} {\bibfield
  {journal} {\bibinfo  {journal} {Phys. Rev. A}\ }\textbf {\bibinfo {volume}
  {67}},\ \bibinfo {pages} {012317} (\bibinfo {year} {2003})}\BibitemShut
  {NoStop}%
\bibitem [{\citenamefont {Torosov}\ and\ \citenamefont
  {Vitanov}(2019{\natexlab{a}})}]{PhysRevA.99.013402}%
  \BibitemOpen
  \bibfield  {author} {\bibinfo {author} {\bibfnamefont {B.~T.}\ \bibnamefont
  {Torosov}}\ and\ \bibinfo {author} {\bibfnamefont {N.~V.}\ \bibnamefont
  {Vitanov}},\ }\bibinfo {title} {Arbitrarily accurate variable rotations on
  the {B}loch sphere by composite pulse sequences},\ \href {\doibase
  10.1103/PhysRevA.99.013402} {\bibfield  {journal} {\bibinfo  {journal} {Phys.
  Rev. A}\ }\textbf {\bibinfo {volume} {99}},\ \bibinfo {pages} {013402}
  (\bibinfo {year} {2019}{\natexlab{a}})}\BibitemShut {NoStop}%
\bibitem [{\citenamefont {Wu}\ \emph {et~al.}(2023)\citenamefont {Wu},
  \citenamefont {Zhang}, \citenamefont {Song}, \citenamefont {Xia},\ and\
  \citenamefont {Shi}}]{PhysRevA.107.023103}%
  \BibitemOpen
  \bibfield  {author} {\bibinfo {author} {\bibfnamefont {H.-N.}\ \bibnamefont
  {Wu}}, \bibinfo {author} {\bibfnamefont {C.}~\bibnamefont {Zhang}}, \bibinfo
  {author} {\bibfnamefont {J.}~\bibnamefont {Song}}, \bibinfo {author}
  {\bibfnamefont {Y.}~\bibnamefont {Xia}}, \ and\ \bibinfo {author}
  {\bibfnamefont {Z.-C.}\ \bibnamefont {Shi}},\ }\bibinfo {title} {Composite
  pulses for optimal robust control in two-level systems},\ \href {\doibase
  10.1103/PhysRevA.107.023103} {\bibfield  {journal} {\bibinfo  {journal}
  {Phys. Rev. A}\ }\textbf {\bibinfo {volume} {107}},\ \bibinfo {pages}
  {023103} (\bibinfo {year} {2023})}\BibitemShut {NoStop}%
\bibitem [{\citenamefont {Kukita}\ \emph {et~al.}(2022)\citenamefont {Kukita},
  \citenamefont {Kiya},\ and\ \citenamefont {Kondo}}]{PhysRevA.106.042613}%
  \BibitemOpen
  \bibfield  {author} {\bibinfo {author} {\bibfnamefont {S.}~\bibnamefont
  {Kukita}}, \bibinfo {author} {\bibfnamefont {H.}~\bibnamefont {Kiya}}, \ and\
  \bibinfo {author} {\bibfnamefont {Y.}~\bibnamefont {Kondo}},\ }\bibinfo
  {title} {General off-resonance-error-robust symmetric composite pulses with
  three elementary operations},\ \href {\doibase 10.1103/PhysRevA.106.042613}
  {\bibfield  {journal} {\bibinfo  {journal} {Phys. Rev. A}\ }\textbf {\bibinfo
  {volume} {106}},\ \bibinfo {pages} {042613} (\bibinfo {year}
  {2022})}\BibitemShut {NoStop}%
\bibitem [{\citenamefont {Zhang}\ \emph {et~al.}(2022)\citenamefont {Zhang},
  \citenamefont {Liu}, \citenamefont {Shi}, \citenamefont {Song}, \citenamefont
  {Xia},\ and\ \citenamefont {Zheng}}]{PhysRevA.105.042414}%
  \BibitemOpen
  \bibfield  {author} {\bibinfo {author} {\bibfnamefont {C.}~\bibnamefont
  {Zhang}}, \bibinfo {author} {\bibfnamefont {Y.}~\bibnamefont {Liu}}, \bibinfo
  {author} {\bibfnamefont {Z.-C.}\ \bibnamefont {Shi}}, \bibinfo {author}
  {\bibfnamefont {J.}~\bibnamefont {Song}}, \bibinfo {author} {\bibfnamefont
  {Y.}~\bibnamefont {Xia}}, \ and\ \bibinfo {author} {\bibfnamefont {S.-B.}\
  \bibnamefont {Zheng}},\ }\bibinfo {title} {Robust population inversion in
  three-level systems by composite pulses},\ \href {\doibase
  10.1103/PhysRevA.105.042414} {\bibfield  {journal} {\bibinfo  {journal}
  {Phys. Rev. A}\ }\textbf {\bibinfo {volume} {105}},\ \bibinfo {pages}
  {042414} (\bibinfo {year} {2022})}\BibitemShut {NoStop}%
\bibitem [{\citenamefont {Cummins}\ \emph {et~al.}(2003)\citenamefont
  {Cummins}, \citenamefont {Llewellyn},\ and\ \citenamefont
  {Jones}}]{PhysRevA.67.042308}%
  \BibitemOpen
  \bibfield  {author} {\bibinfo {author} {\bibfnamefont {H.~K.}\ \bibnamefont
  {Cummins}}, \bibinfo {author} {\bibfnamefont {G.}~\bibnamefont {Llewellyn}},
  \ and\ \bibinfo {author} {\bibfnamefont {J.~A.}\ \bibnamefont {Jones}},\
  }\bibinfo {title} {Tackling systematic errors in quantum logic gates with
  composite rotations},\ \href {\doibase 10.1103/PhysRevA.67.042308} {\bibfield
   {journal} {\bibinfo  {journal} {Phys. Rev. A}\ }\textbf {\bibinfo {volume}
  {67}},\ \bibinfo {pages} {042308} (\bibinfo {year} {2003})}\BibitemShut
  {NoStop}%
\bibitem [{\citenamefont {Ivanov}\ and\ \citenamefont
  {Vitanov}(2015)}]{PhysRevA.92.022333}%
  \BibitemOpen
  \bibfield  {author} {\bibinfo {author} {\bibfnamefont {S.~S.}\ \bibnamefont
  {Ivanov}}\ and\ \bibinfo {author} {\bibfnamefont {N.~V.}\ \bibnamefont
  {Vitanov}},\ }\bibinfo {title} {Composite two-qubit gates},\ \href {\doibase
  10.1103/PhysRevA.92.022333} {\bibfield  {journal} {\bibinfo  {journal} {Phys.
  Rev. A}\ }\textbf {\bibinfo {volume} {92}},\ \bibinfo {pages} {022333}
  (\bibinfo {year} {2015})}\BibitemShut {NoStop}%
\bibitem [{\citenamefont {Ichikawa}\ \emph {et~al.}(2011)\citenamefont
  {Ichikawa}, \citenamefont {Bando}, \citenamefont {Kondo},\ and\ \citenamefont
  {Nakahara}}]{PhysRevA.84.062311}%
  \BibitemOpen
  \bibfield  {author} {\bibinfo {author} {\bibfnamefont {T.}~\bibnamefont
  {Ichikawa}}, \bibinfo {author} {\bibfnamefont {M.}~\bibnamefont {Bando}},
  \bibinfo {author} {\bibfnamefont {Y.}~\bibnamefont {Kondo}}, \ and\ \bibinfo
  {author} {\bibfnamefont {M.}~\bibnamefont {Nakahara}},\ }\bibinfo {title}
  {Designing robust unitary gates: application to concatenated composite
  pulses},\ \href {\doibase 10.1103/PhysRevA.84.062311} {\bibfield  {journal}
  {\bibinfo  {journal} {Phys. Rev. A}\ }\textbf {\bibinfo {volume} {84}},\
  \bibinfo {pages} {062311} (\bibinfo {year} {2011})}\BibitemShut {NoStop}%
\bibitem [{\citenamefont {Cohen}\ \emph {et~al.}(2016)\citenamefont {Cohen},
  \citenamefont {Rotem},\ and\ \citenamefont {Retzker}}]{PhysRevA.93.032340}%
  \BibitemOpen
  \bibfield  {author} {\bibinfo {author} {\bibfnamefont {I.}~\bibnamefont
  {Cohen}}, \bibinfo {author} {\bibfnamefont {A.}~\bibnamefont {Rotem}}, \ and\
  \bibinfo {author} {\bibfnamefont {A.}~\bibnamefont {Retzker}},\ }\bibinfo
  {title} {Refocusing two-qubit-gate noise for trapped ions by composite
  pulses},\ \href {\doibase 10.1103/PhysRevA.93.032340} {\bibfield  {journal}
  {\bibinfo  {journal} {Phys. Rev. A}\ }\textbf {\bibinfo {volume} {93}},\
  \bibinfo {pages} {032340} (\bibinfo {year} {2016})}\BibitemShut {NoStop}%
\bibitem [{\citenamefont {Torosov}\ and\ \citenamefont
  {Vitanov}(2014)}]{PhysRevA.90.012341}%
  \BibitemOpen
  \bibfield  {author} {\bibinfo {author} {\bibfnamefont {B.~T.}\ \bibnamefont
  {Torosov}}\ and\ \bibinfo {author} {\bibfnamefont {N.~V.}\ \bibnamefont
  {Vitanov}},\ }\bibinfo {title} {High-fidelity error-resilient composite phase
  gates},\ \href {\doibase 10.1103/PhysRevA.90.012341} {\bibfield  {journal}
  {\bibinfo  {journal} {Phys. Rev. A}\ }\textbf {\bibinfo {volume} {90}},\
  \bibinfo {pages} {012341} (\bibinfo {year} {2014})}\BibitemShut {NoStop}%
\bibitem [{\citenamefont {Mount}\ \emph {et~al.}(2015)\citenamefont {Mount},
  \citenamefont {Kabytayev}, \citenamefont {Crain}, \citenamefont {Harper},
  \citenamefont {Baek}, \citenamefont {Vrijsen}, \citenamefont {Flammia},
  \citenamefont {Brown}, \citenamefont {Maunz},\ and\ \citenamefont
  {Kim}}]{PhysRevA.92.060301}%
  \BibitemOpen
  \bibfield  {author} {\bibinfo {author} {\bibfnamefont {E.}~\bibnamefont
  {Mount}}, \bibinfo {author} {\bibfnamefont {C.}~\bibnamefont {Kabytayev}},
  \bibinfo {author} {\bibfnamefont {S.}~\bibnamefont {Crain}}, \bibinfo
  {author} {\bibfnamefont {R.}~\bibnamefont {Harper}}, \bibinfo {author}
  {\bibfnamefont {S.-Y.}\ \bibnamefont {Baek}}, \bibinfo {author}
  {\bibfnamefont {G.}~\bibnamefont {Vrijsen}}, \bibinfo {author} {\bibfnamefont
  {S.~T.}\ \bibnamefont {Flammia}}, \bibinfo {author} {\bibfnamefont {K.~R.}\
  \bibnamefont {Brown}}, \bibinfo {author} {\bibfnamefont {P.}~\bibnamefont
  {Maunz}}, \ and\ \bibinfo {author} {\bibfnamefont {J.}~\bibnamefont {Kim}},\
  }\bibinfo {title} {Error compensation of single-qubit gates in a
  surface-electrode ion trap using composite pulses},\ \href {\doibase
  10.1103/PhysRevA.92.060301} {\bibfield  {journal} {\bibinfo  {journal} {Phys.
  Rev. A}\ }\textbf {\bibinfo {volume} {92}},\ \bibinfo {pages} {060301(R)}
  (\bibinfo {year} {2015})}\BibitemShut {NoStop}%
\bibitem [{\citenamefont {Torosov}\ and\ \citenamefont
  {Vitanov}(2020)}]{PhysRevResearch.2.043194}%
  \BibitemOpen
  \bibfield  {author} {\bibinfo {author} {\bibfnamefont {B.~T.}\ \bibnamefont
  {Torosov}}\ and\ \bibinfo {author} {\bibfnamefont {N.~V.}\ \bibnamefont
  {Vitanov}},\ }\bibinfo {title} {High-fidelity composite quantum gates for
  {R}aman qubits},\ \href {\doibase 10.1103/PhysRevResearch.2.043194}
  {\bibfield  {journal} {\bibinfo  {journal} {Phys. Rev. Research}\ }\textbf
  {\bibinfo {volume} {2}},\ \bibinfo {pages} {043194} (\bibinfo {year}
  {2020})}\BibitemShut {NoStop}%
\bibitem [{\citenamefont {Shi}\ \emph {et~al.}(2021)\citenamefont {Shi},
  \citenamefont {Wu}, \citenamefont {Shen}, \citenamefont {Song}, \citenamefont
  {Xia}, \citenamefont {Yi},\ and\ \citenamefont
  {Zheng}}]{PhysRevA.103.052612}%
  \BibitemOpen
  \bibfield  {author} {\bibinfo {author} {\bibfnamefont {Z.-C.}\ \bibnamefont
  {Shi}}, \bibinfo {author} {\bibfnamefont {H.-N.}\ \bibnamefont {Wu}},
  \bibinfo {author} {\bibfnamefont {L.-T.}\ \bibnamefont {Shen}}, \bibinfo
  {author} {\bibfnamefont {J.}~\bibnamefont {Song}}, \bibinfo {author}
  {\bibfnamefont {Y.}~\bibnamefont {Xia}}, \bibinfo {author} {\bibfnamefont
  {X.~X.}\ \bibnamefont {Yi}}, \ and\ \bibinfo {author} {\bibfnamefont {S.-B.}\
  \bibnamefont {Zheng}},\ }\bibinfo {title} {Robust single-qubit gates by
  composite pulses in three-level systems},\ \href {\doibase
  10.1103/PhysRevA.103.052612} {\bibfield  {journal} {\bibinfo  {journal}
  {Phys. Rev. A}\ }\textbf {\bibinfo {volume} {103}},\ \bibinfo {pages}
  {052612} (\bibinfo {year} {2021})}\BibitemShut {NoStop}%
\bibitem [{\citenamefont {Xiao}\ and\ \citenamefont
  {Jones}(2006)}]{PhysRevA.73.032334}%
  \BibitemOpen
  \bibfield  {author} {\bibinfo {author} {\bibfnamefont {L.}~\bibnamefont
  {Xiao}}\ and\ \bibinfo {author} {\bibfnamefont {J.~A.}\ \bibnamefont
  {Jones}},\ }\bibinfo {title} {Robust logic gates and realistic quantum
  computation},\ \href {\doibase 10.1103/PhysRevA.73.032334} {\bibfield
  {journal} {\bibinfo  {journal} {Phys. Rev. A}\ }\textbf {\bibinfo {volume}
  {73}},\ \bibinfo {pages} {032334} (\bibinfo {year} {2006})}\BibitemShut
  {NoStop}%
\bibitem [{\citenamefont {H$\ddot{\mathrm{a}}$ffner}\ \emph
  {et~al.}(2008)\citenamefont {H$\ddot{\mathrm{a}}$ffner}, \citenamefont
  {Roos},\ and\ \citenamefont {Blatt}}]{HAFFNER2008155}%
  \BibitemOpen
  \bibfield  {author} {\bibinfo {author} {\bibfnamefont {H.}~\bibnamefont
  {H$\ddot{\mathrm{a}}$ffner}}, \bibinfo {author} {\bibfnamefont
  {C.}~\bibnamefont {Roos}}, \ and\ \bibinfo {author} {\bibfnamefont
  {R.}~\bibnamefont {Blatt}},\ }\bibinfo {title} {Quantum computing with
  trapped ions},\ \href {\doibase
  https://doi.org/10.1016/j.physrep.2008.09.003} {\bibfield  {journal}
  {\bibinfo  {journal} {Phys. Rep.}\ }\textbf {\bibinfo {volume} {469}},\
  \bibinfo {pages} {155} (\bibinfo {year} {2008})}\BibitemShut {NoStop}%
\bibitem [{\citenamefont {Joo}\ \emph {et~al.}(2006)\citenamefont {Joo},
  \citenamefont {Lim}, \citenamefont {Beige},\ and\ \citenamefont
  {Knight}}]{PhysRevA.74.042344}%
  \BibitemOpen
  \bibfield  {author} {\bibinfo {author} {\bibfnamefont {J.}~\bibnamefont
  {Joo}}, \bibinfo {author} {\bibfnamefont {Y.~L.}\ \bibnamefont {Lim}},
  \bibinfo {author} {\bibfnamefont {A.}~\bibnamefont {Beige}}, \ and\ \bibinfo
  {author} {\bibfnamefont {P.~L.}\ \bibnamefont {Knight}},\ }\bibinfo {title}
  {Single-qubit rotations in two-dimensional optical lattices with multiqubit
  addressing},\ \href {\doibase 10.1103/PhysRevA.74.042344} {\bibfield
  {journal} {\bibinfo  {journal} {Phys. Rev. A}\ }\textbf {\bibinfo {volume}
  {74}},\ \bibinfo {pages} {042344} (\bibinfo {year} {2006})}\BibitemShut
  {NoStop}%
\bibitem [{\citenamefont {LeBlanc}\ and\ \citenamefont
  {Thywissen}(2007)}]{PhysRevA.75.053612}%
  \BibitemOpen
  \bibfield  {author} {\bibinfo {author} {\bibfnamefont {L.~J.}\ \bibnamefont
  {LeBlanc}}\ and\ \bibinfo {author} {\bibfnamefont {J.~H.}\ \bibnamefont
  {Thywissen}},\ }\bibinfo {title} {Species-specific optical lattices},\ \href
  {\doibase 10.1103/PhysRevA.75.053612} {\bibfield  {journal} {\bibinfo
  {journal} {Phys. Rev. A}\ }\textbf {\bibinfo {volume} {75}},\ \bibinfo
  {pages} {053612} (\bibinfo {year} {2007})}\BibitemShut {NoStop}%
\bibitem [{\citenamefont {Cho}(2007)}]{PhysRevLett.99.020502}%
  \BibitemOpen
  \bibfield  {author} {\bibinfo {author} {\bibfnamefont {J.}~\bibnamefont
  {Cho}},\ }\bibinfo {title} {Addressing Individual Atoms in Optical Lattices
  with Standing-Wave Driving Fields},\ \href {\doibase
  10.1103/PhysRevLett.99.020502} {\bibfield  {journal} {\bibinfo  {journal}
  {Phys. Rev. Lett.}\ }\textbf {\bibinfo {volume} {99}},\ \bibinfo {pages}
  {020502} (\bibinfo {year} {2007})}\BibitemShut {NoStop}%
\bibitem [{\citenamefont {Tycko}\ and\ \citenamefont
  {Pines}(1984)}]{TYCKO1984462}%
  \BibitemOpen
  \bibfield  {author} {\bibinfo {author} {\bibfnamefont {R.}~\bibnamefont
  {Tycko}}\ and\ \bibinfo {author} {\bibfnamefont {A.}~\bibnamefont {Pines}},\
  }\bibinfo {title} {Iterative schemes for broad-band and narrow-band
  population inversion in NMR},\ \href {\doibase
  https://doi.org/10.1016/0009-2614(84)85541-4} {\bibfield  {journal} {\bibinfo
   {journal} {Chem. Phys. Lett.}\ }\textbf {\bibinfo {volume} {111}},\ \bibinfo
  {pages} {462} (\bibinfo {year} {1984})}\BibitemShut {NoStop}%
\bibitem [{\citenamefont {Ivanov}\ and\ \citenamefont
  {Vitanov}(2011)}]{Ivanov2011}%
  \BibitemOpen
  \bibfield  {author} {\bibinfo {author} {\bibfnamefont {S.~S.}\ \bibnamefont
  {Ivanov}}\ and\ \bibinfo {author} {\bibfnamefont {N.~V.}\ \bibnamefont
  {Vitanov}},\ }\bibinfo {title} {High-fidelity local addressing of trapped
  ions and atoms by composite sequences of laser pulses},\ \href {\doibase
  10.1364/ol.36.001275} {\bibfield  {journal} {\bibinfo  {journal} {Opt.
  Lett.}\ }\textbf {\bibinfo {volume} {36}},\ \bibinfo {pages} {1275} (\bibinfo
  {year} {2011})}\BibitemShut {NoStop}%
\bibitem [{\citenamefont {Shaka}\ and\ \citenamefont
  {Freeman}(1984)}]{SHAKA1984169}%
  \BibitemOpen
  \bibfield  {author} {\bibinfo {author} {\bibfnamefont {A.}~\bibnamefont
  {Shaka}}\ and\ \bibinfo {author} {\bibfnamefont {R.}~\bibnamefont
  {Freeman}},\ }\bibinfo {title} {Spatially selective radiofrequency pulses},\
  \href {\doibase https://doi.org/10.1016/0022-2364(84)90297-X} {\bibfield
  {journal} {\bibinfo  {journal} {J. Magn. Reson.}\ }\textbf {\bibinfo {volume}
  {59}},\ \bibinfo {pages} {169} (\bibinfo {year} {1984})}\BibitemShut
  {NoStop}%
\bibitem [{\citenamefont {Torosov}\ \emph {et~al.}(2020)\citenamefont
  {Torosov}, \citenamefont {Ivanov},\ and\ \citenamefont
  {Vitanov}}]{PhysRevA.102.013105}%
  \BibitemOpen
  \bibfield  {author} {\bibinfo {author} {\bibfnamefont {B.~T.}\ \bibnamefont
  {Torosov}}, \bibinfo {author} {\bibfnamefont {S.~S.}\ \bibnamefont {Ivanov}},
  \ and\ \bibinfo {author} {\bibfnamefont {N.~V.}\ \bibnamefont {Vitanov}},\
  }\bibinfo {title} {Narrowband and passband composite pulses for variable
  rotations},\ \href {\doibase 10.1103/PhysRevA.102.013105} {\bibfield
  {journal} {\bibinfo  {journal} {Phys. Rev. A}\ }\textbf {\bibinfo {volume}
  {102}},\ \bibinfo {pages} {013105} (\bibinfo {year} {2020})}\BibitemShut
  {NoStop}%
\bibitem [{\citenamefont {Vitanov}(2011)}]{PhysRevA.84.065404}%
  \BibitemOpen
  \bibfield  {author} {\bibinfo {author} {\bibfnamefont {N.~V.}\ \bibnamefont
  {Vitanov}},\ }\bibinfo {title} {Arbitrarily accurate narrowband composite
  pulse sequences},\ \href {\doibase 10.1103/PhysRevA.84.065404} {\bibfield
  {journal} {\bibinfo  {journal} {Phys. Rev. A}\ }\textbf {\bibinfo {volume}
  {84}},\ \bibinfo {pages} {065404} (\bibinfo {year} {2011})}\BibitemShut
  {NoStop}%
\bibitem [{\citenamefont {Torosov}\ and\ \citenamefont
  {Vitanov}(2023)}]{PhysRevA.107.032618}%
  \BibitemOpen
  \bibfield  {author} {\bibinfo {author} {\bibfnamefont {B.~T.}\ \bibnamefont
  {Torosov}}\ and\ \bibinfo {author} {\bibfnamefont {N.~V.}\ \bibnamefont
  {Vitanov}},\ }\bibinfo {title} {Narrowband composite two-qubit gates for
  crosstalk suppression},\ \href {\doibase 10.1103/PhysRevA.107.032618}
  {\bibfield  {journal} {\bibinfo  {journal} {Phys. Rev. A}\ }\textbf {\bibinfo
  {volume} {107}},\ \bibinfo {pages} {032618} (\bibinfo {year}
  {2023})}\BibitemShut {NoStop}%
\bibitem [{\citenamefont {Torosov}\ and\ \citenamefont
  {Vitanov}(2011)}]{PhysRevA.83.053420}%
  \BibitemOpen
  \bibfield  {author} {\bibinfo {author} {\bibfnamefont {B.~T.}\ \bibnamefont
  {Torosov}}\ and\ \bibinfo {author} {\bibfnamefont {N.~V.}\ \bibnamefont
  {Vitanov}},\ }\bibinfo {title} {Smooth composite pulses for high-fidelity
  quantum information processing},\ \href {\doibase 10.1103/PhysRevA.83.053420}
  {\bibfield  {journal} {\bibinfo  {journal} {Phys. Rev. A}\ }\textbf {\bibinfo
  {volume} {83}},\ \bibinfo {pages} {053420} (\bibinfo {year}
  {2011})}\BibitemShut {NoStop}%
\bibitem [{\citenamefont {Kyoseva}\ and\ \citenamefont
  {Vitanov}(2013)}]{PhysRevA.88.063410}%
  \BibitemOpen
  \bibfield  {author} {\bibinfo {author} {\bibfnamefont {E.}~\bibnamefont
  {Kyoseva}}\ and\ \bibinfo {author} {\bibfnamefont {N.~V.}\ \bibnamefont
  {Vitanov}},\ }\bibinfo {title} {Arbitrarily accurate passband composite
  pulses for dynamical suppression of amplitude noise},\ \href {\doibase
  10.1103/PhysRevA.88.063410} {\bibfield  {journal} {\bibinfo  {journal} {Phys.
  Rev. A}\ }\textbf {\bibinfo {volume} {88}},\ \bibinfo {pages} {063410}
  (\bibinfo {year} {2013})}\BibitemShut {NoStop}%
\bibitem [{\citenamefont {Cho}\ \emph {et~al.}(1986)\citenamefont {Cho},
  \citenamefont {Tycko}, \citenamefont {Pines},\ and\ \citenamefont
  {Guckenheimer}}]{PhysRevLett.56.1905}%
  \BibitemOpen
  \bibfield  {author} {\bibinfo {author} {\bibfnamefont {H.~M.}\ \bibnamefont
  {Cho}}, \bibinfo {author} {\bibfnamefont {R.}~\bibnamefont {Tycko}}, \bibinfo
  {author} {\bibfnamefont {A.}~\bibnamefont {Pines}}, \ and\ \bibinfo {author}
  {\bibfnamefont {J.}~\bibnamefont {Guckenheimer}},\ }\bibinfo {title}
  {Iterative maps for bistable excitation of two-level systems},\ \href
  {\doibase 10.1103/PhysRevLett.56.1905} {\bibfield  {journal} {\bibinfo
  {journal} {Phys. Rev. Lett.}\ }\textbf {\bibinfo {volume} {56}},\ \bibinfo
  {pages} {1905} (\bibinfo {year} {1986})}\BibitemShut {NoStop}%
\bibitem [{\citenamefont {Merrill}\ \emph {et~al.}(2014)\citenamefont
  {Merrill}, \citenamefont {Doret}, \citenamefont {Vittorini}, \citenamefont
  {Addison},\ and\ \citenamefont {Brown}}]{PhysRevA.90.040301}%
  \BibitemOpen
  \bibfield  {author} {\bibinfo {author} {\bibfnamefont {J.~T.}\ \bibnamefont
  {Merrill}}, \bibinfo {author} {\bibfnamefont {S.~C.}\ \bibnamefont {Doret}},
  \bibinfo {author} {\bibfnamefont {G.}~\bibnamefont {Vittorini}}, \bibinfo
  {author} {\bibfnamefont {J.~P.}\ \bibnamefont {Addison}}, \ and\ \bibinfo
  {author} {\bibfnamefont {K.~R.}\ \bibnamefont {Brown}},\ }\bibinfo {title}
  {Transformed composite sequences for improved qubit addressing},\ \href
  {\doibase 10.1103/PhysRevA.90.040301} {\bibfield  {journal} {\bibinfo
  {journal} {Phys. Rev. A}\ }\textbf {\bibinfo {volume} {90}},\ \bibinfo
  {pages} {040301(R)} (\bibinfo {year} {2014})}\BibitemShut {NoStop}%
\bibitem [{\citenamefont {Genov}\ \emph {et~al.}(2011)\citenamefont {Genov},
  \citenamefont {Torosov},\ and\ \citenamefont {Vitanov}}]{PhysRevA.84.063413}%
  \BibitemOpen
  \bibfield  {author} {\bibinfo {author} {\bibfnamefont {G.~T.}\ \bibnamefont
  {Genov}}, \bibinfo {author} {\bibfnamefont {B.~T.}\ \bibnamefont {Torosov}},
  \ and\ \bibinfo {author} {\bibfnamefont {N.~V.}\ \bibnamefont {Vitanov}},\
  }\bibinfo {title} {Optimized control of multistate quantum systems by
  composite pulse sequences},\ \href {\doibase 10.1103/PhysRevA.84.063413}
  {\bibfield  {journal} {\bibinfo  {journal} {Phys. Rev. A}\ }\textbf {\bibinfo
  {volume} {84}},\ \bibinfo {pages} {063413} (\bibinfo {year}
  {2011})}\BibitemShut {NoStop}%
\bibitem [{\citenamefont {Husain}\ \emph {et~al.}(2013)\citenamefont {Husain},
  \citenamefont {Kawamura},\ and\ \citenamefont {Jones}}]{Husain2013}%
  \BibitemOpen
  \bibfield  {author} {\bibinfo {author} {\bibfnamefont {S.}~\bibnamefont
  {Husain}}, \bibinfo {author} {\bibfnamefont {M.}~\bibnamefont {Kawamura}}, \
  and\ \bibinfo {author} {\bibfnamefont {J.~A.}\ \bibnamefont {Jones}},\
  }\bibinfo {title} {Further analysis of some symmetric and antisymmetric
  composite pulses for tackling pulse strength errors},\ \href {\doibase
  10.1016/j.jmr.2013.02.007} {\bibfield  {journal} {\bibinfo  {journal} {J.
  Magn. Reson.}\ }\textbf {\bibinfo {volume} {230}},\ \bibinfo {pages} {145}
  (\bibinfo {year} {2013})}\BibitemShut {NoStop}%
\bibitem [{\citenamefont {Morris}\ and\ \citenamefont
  {Shore}(1983)}]{PhysRevA.27.906}%
  \BibitemOpen
  \bibfield  {author} {\bibinfo {author} {\bibfnamefont {J.~R.}\ \bibnamefont
  {Morris}}\ and\ \bibinfo {author} {\bibfnamefont {B.~W.}\ \bibnamefont
  {Shore}},\ }\bibinfo {title} {Reduction of degenerate two-level excitation to
  independent two-state systems},\ \href {\doibase 10.1103/PhysRevA.27.906}
  {\bibfield  {journal} {\bibinfo  {journal} {Phys. Rev. A}\ }\textbf {\bibinfo
  {volume} {27}},\ \bibinfo {pages} {906} (\bibinfo {year} {1983})}\BibitemShut
  {NoStop}%
\bibitem [{\citenamefont {Majorana}(1932)}]{Majorana1932}%
  \BibitemOpen
  \bibfield  {author} {\bibinfo {author} {\bibfnamefont {E.}~\bibnamefont
  {Majorana}},\ }\bibinfo {title} {Atomi orientati in campo magnetico
  variabile},\ \href {\doibase 10.1007/bf02960953} {\bibfield  {journal}
  {\bibinfo  {journal} {Nuovo Cimento}\ }\textbf {\bibinfo {volume} {9}},\
  \bibinfo {pages} {43} (\bibinfo {year} {1932})}\BibitemShut {NoStop}%
\bibitem [{\citenamefont {Zhang}\ \emph {et~al.}(2017)\citenamefont {Zhang},
  \citenamefont {Zhou}, \citenamefont {Zhou}, \citenamefont {Guo},\ and\
  \citenamefont {Zhou}}]{PhysRevLett.118.083604}%
  \BibitemOpen
  \bibfield  {author} {\bibinfo {author} {\bibfnamefont {Y.-C.}\ \bibnamefont
  {Zhang}}, \bibinfo {author} {\bibfnamefont {X.-F.}\ \bibnamefont {Zhou}},
  \bibinfo {author} {\bibfnamefont {X.}~\bibnamefont {Zhou}}, \bibinfo {author}
  {\bibfnamefont {G.-C.}\ \bibnamefont {Guo}}, \ and\ \bibinfo {author}
  {\bibfnamefont {Z.-W.}\ \bibnamefont {Zhou}},\ }\bibinfo {title}
  {Cavity-Assisted Single-Mode and Two-Mode Spin-Squeezed States via
  Phase-Locked Atom-Photon Coupling},\ \href {\doibase
  10.1103/PhysRevLett.118.083604} {\bibfield  {journal} {\bibinfo  {journal}
  {Phys. Rev. Lett.}\ }\textbf {\bibinfo {volume} {118}},\ \bibinfo {pages}
  {083604} (\bibinfo {year} {2017})}\BibitemShut {NoStop}%
\bibitem [{\citenamefont {Wollenhaupt}\ \emph {et~al.}(2005)\citenamefont
  {Wollenhaupt}, \citenamefont {Engel},\ and\ \citenamefont
  {Baumert}}]{Wollenhaupt2005}%
  \BibitemOpen
  \bibfield  {author} {\bibinfo {author} {\bibfnamefont {M.}~\bibnamefont
  {Wollenhaupt}}, \bibinfo {author} {\bibfnamefont {V.}~\bibnamefont {Engel}},
  \ and\ \bibinfo {author} {\bibfnamefont {T.}~\bibnamefont {Baumert}},\
  }\bibinfo {title} {Femtosecond laser photoelectron spectroscopy on atoms and
  small molecules: Prototype studies in quantum control},\ \href {\doibase
  10.1146/annurev.physchem.56.092503.141315} {\bibfield  {journal} {\bibinfo
  {journal} {Annu. Rev. Phys. Chem.}\ }\textbf {\bibinfo {volume} {56}},\
  \bibinfo {pages} {25} (\bibinfo {year} {2005})}\BibitemShut {NoStop}%
\bibitem [{\citenamefont {Cho}\ \emph {et~al.}(2008)\citenamefont {Cho},
  \citenamefont {Angelakis},\ and\ \citenamefont
  {Bose}}]{PhysRevLett.101.246809}%
  \BibitemOpen
  \bibfield  {author} {\bibinfo {author} {\bibfnamefont {J.}~\bibnamefont
  {Cho}}, \bibinfo {author} {\bibfnamefont {D.~G.}\ \bibnamefont {Angelakis}},
  \ and\ \bibinfo {author} {\bibfnamefont {S.}~\bibnamefont {Bose}},\ }\bibinfo
  {title} {Fractional Quantum Hall State in Coupled Cavities},\ \href {\doibase
  10.1103/PhysRevLett.101.246809} {\bibfield  {journal} {\bibinfo  {journal}
  {Phys. Rev. Lett.}\ }\textbf {\bibinfo {volume} {101}},\ \bibinfo {pages}
  {246809} (\bibinfo {year} {2008})}\BibitemShut {NoStop}%
\bibitem [{\citenamefont {Ye}\ \emph {et~al.}(2019)\citenamefont {Ye},
  \citenamefont {Zhang}, \citenamefont {Chen},\ and\ \citenamefont
  {Li}}]{PhysRevA.100.043403}%
  \BibitemOpen
  \bibfield  {author} {\bibinfo {author} {\bibfnamefont {C.}~\bibnamefont
  {Ye}}, \bibinfo {author} {\bibfnamefont {Q.}~\bibnamefont {Zhang}}, \bibinfo
  {author} {\bibfnamefont {Y.-Y.}\ \bibnamefont {Chen}}, \ and\ \bibinfo
  {author} {\bibfnamefont {Y.}~\bibnamefont {Li}},\ }\bibinfo {title}
  {Effective two-level models for highly efficient inner-state
  enantioseparation based on cyclic three-level systems of chiral molecules},\
  \href {\doibase 10.1103/PhysRevA.100.043403} {\bibfield  {journal} {\bibinfo
  {journal} {Phys. Rev. A}\ }\textbf {\bibinfo {volume} {100}},\ \bibinfo
  {pages} {043403} (\bibinfo {year} {2019})}\BibitemShut {NoStop}%
\bibitem [{\citenamefont {Kestner}\ \emph {et~al.}(2013)\citenamefont
  {Kestner}, \citenamefont {Wang}, \citenamefont {Bishop}, \citenamefont
  {Barnes},\ and\ \citenamefont {Das~Sarma}}]{PhysRevLett.110.140502}%
  \BibitemOpen
  \bibfield  {author} {\bibinfo {author} {\bibfnamefont {J.~P.}\ \bibnamefont
  {Kestner}}, \bibinfo {author} {\bibfnamefont {X.}~\bibnamefont {Wang}},
  \bibinfo {author} {\bibfnamefont {L.~S.}\ \bibnamefont {Bishop}}, \bibinfo
  {author} {\bibfnamefont {E.}~\bibnamefont {Barnes}}, \ and\ \bibinfo {author}
  {\bibfnamefont {S.}~\bibnamefont {Das~Sarma}},\ }\bibinfo {title}
  {Noise-resistant control for a spin qubit array},\ \href {\doibase
  10.1103/PhysRevLett.110.140502} {\bibfield  {journal} {\bibinfo  {journal}
  {Phys. Rev. Lett.}\ }\textbf {\bibinfo {volume} {110}},\ \bibinfo {pages}
  {140502} (\bibinfo {year} {2013})}\BibitemShut {NoStop}%
\bibitem [{\citenamefont {Ghosh}\ \emph {et~al.}(2017)\citenamefont {Ghosh},
  \citenamefont {Coppersmith},\ and\ \citenamefont
  {Friesen}}]{PhysRevB.95.241307}%
  \BibitemOpen
  \bibfield  {author} {\bibinfo {author} {\bibfnamefont {J.}~\bibnamefont
  {Ghosh}}, \bibinfo {author} {\bibfnamefont {S.~N.}\ \bibnamefont
  {Coppersmith}}, \ and\ \bibinfo {author} {\bibfnamefont {M.}~\bibnamefont
  {Friesen}},\ }\bibinfo {title} {Pulse sequences for suppressing leakage in
  single-qubit gate operations},\ \href {\doibase 10.1103/PhysRevB.95.241307}
  {\bibfield  {journal} {\bibinfo  {journal} {Phys. Rev. B}\ }\textbf {\bibinfo
  {volume} {95}},\ \bibinfo {pages} {241307(R)} (\bibinfo {year}
  {2017})}\BibitemShut {NoStop}%
\bibitem [{\citenamefont {Torosov}\ \emph {et~al.}(2021)\citenamefont
  {Torosov}, \citenamefont {Shore},\ and\ \citenamefont
  {Vitanov}}]{PhysRevA.103.033110}%
  \BibitemOpen
  \bibfield  {author} {\bibinfo {author} {\bibfnamefont {B.~T.}\ \bibnamefont
  {Torosov}}, \bibinfo {author} {\bibfnamefont {B.~W.}\ \bibnamefont {Shore}},
  \ and\ \bibinfo {author} {\bibfnamefont {N.~V.}\ \bibnamefont {Vitanov}},\
  }\bibinfo {title} {Coherent control techniques for two-state quantum systems:
  a comparative study},\ \href {\doibase 10.1103/PhysRevA.103.033110}
  {\bibfield  {journal} {\bibinfo  {journal} {Phys. Rev. A}\ }\textbf {\bibinfo
  {volume} {103}},\ \bibinfo {pages} {033110} (\bibinfo {year}
  {2021})}\BibitemShut {NoStop}%
\bibitem [{\citenamefont {Torosov}\ and\ \citenamefont
  {Vitanov}(2019{\natexlab{b}})}]{PhysRevA.100.023410}%
  \BibitemOpen
  \bibfield  {author} {\bibinfo {author} {\bibfnamefont {B.~T.}\ \bibnamefont
  {Torosov}}\ and\ \bibinfo {author} {\bibfnamefont {N.~V.}\ \bibnamefont
  {Vitanov}},\ }\bibinfo {title} {Composite pulses with errant phases},\ \href
  {\doibase 10.1103/PhysRevA.100.023410} {\bibfield  {journal} {\bibinfo
  {journal} {Phys. Rev. A}\ }\textbf {\bibinfo {volume} {100}},\ \bibinfo
  {pages} {023410} (\bibinfo {year} {2019}{\natexlab{b}})}\BibitemShut
  {NoStop}%
\bibitem [{\citenamefont {Xiang}\ \emph {et~al.}(2013)\citenamefont {Xiang},
  \citenamefont {Ashhab}, \citenamefont {You},\ and\ \citenamefont
  {Nori}}]{RevModPhys.85.623}%
  \BibitemOpen
  \bibfield  {author} {\bibinfo {author} {\bibfnamefont {Z.-L.}\ \bibnamefont
  {Xiang}}, \bibinfo {author} {\bibfnamefont {S.}~\bibnamefont {Ashhab}},
  \bibinfo {author} {\bibfnamefont {J.~Q.}\ \bibnamefont {You}}, \ and\
  \bibinfo {author} {\bibfnamefont {F.}~\bibnamefont {Nori}},\ }\bibinfo
  {title} {Hybrid quantum circuits: Superconducting circuits interacting with
  other quantum systems},\ \href {\doibase 10.1103/RevModPhys.85.623}
  {\bibfield  {journal} {\bibinfo  {journal} {Rev. Mod. Phys.}\ }\textbf
  {\bibinfo {volume} {85}},\ \bibinfo {pages} {623} (\bibinfo {year}
  {2013})}\BibitemShut {NoStop}%
\bibitem [{\citenamefont {Niemczyk}\ \emph {et~al.}(2010)\citenamefont
  {Niemczyk}, \citenamefont {Deppe}, \citenamefont {Huebl}, \citenamefont
  {Menzel}, \citenamefont {Hocke}, \citenamefont {Schwarz}, \citenamefont
  {Garcia-Ripoll}, \citenamefont {Zueco}, \citenamefont {H\"{u}mmer},
  \citenamefont {Solano}, \citenamefont {Marx},\ and\ \citenamefont
  {Gross}}]{Niemczyk2010}%
  \BibitemOpen
  \bibfield  {author} {\bibinfo {author} {\bibfnamefont {T.}~\bibnamefont
  {Niemczyk}}, \bibinfo {author} {\bibfnamefont {F.}~\bibnamefont {Deppe}},
  \bibinfo {author} {\bibfnamefont {H.}~\bibnamefont {Huebl}}, \bibinfo
  {author} {\bibfnamefont {E.~P.}\ \bibnamefont {Menzel}}, \bibinfo {author}
  {\bibfnamefont {F.}~\bibnamefont {Hocke}}, \bibinfo {author} {\bibfnamefont
  {M.~J.}\ \bibnamefont {Schwarz}}, \bibinfo {author} {\bibfnamefont {J.~J.}\
  \bibnamefont {Garcia-Ripoll}}, \bibinfo {author} {\bibfnamefont
  {D.}~\bibnamefont {Zueco}}, \bibinfo {author} {\bibfnamefont
  {T.}~\bibnamefont {H\"{u}mmer}}, \bibinfo {author} {\bibfnamefont
  {E.}~\bibnamefont {Solano}}, \bibinfo {author} {\bibfnamefont
  {A.}~\bibnamefont {Marx}}, \ and\ \bibinfo {author} {\bibfnamefont
  {R.}~\bibnamefont {Gross}},\ }\bibinfo {title} {Circuit quantum
  electrodynamics in the ultrastrong-coupling regime},\ \href {\doibase
  10.1038/nphys1730} {\bibfield  {journal} {\bibinfo  {journal} {Nat. Phys.}\
  }\textbf {\bibinfo {volume} {6}},\ \bibinfo {pages} {772} (\bibinfo {year}
  {2010})}\BibitemShut {NoStop}%
\bibitem [{\citenamefont {Jaako}\ \emph {et~al.}(2019)\citenamefont {Jaako},
  \citenamefont {Garc\'{\i}a-Ripoll},\ and\ \citenamefont
  {Rabl}}]{PhysRevA.100.043815}%
  \BibitemOpen
  \bibfield  {author} {\bibinfo {author} {\bibfnamefont {T.}~\bibnamefont
  {Jaako}}, \bibinfo {author} {\bibfnamefont {J.~J.}\ \bibnamefont
  {Garc\'{\i}a-Ripoll}}, \ and\ \bibinfo {author} {\bibfnamefont
  {P.}~\bibnamefont {Rabl}},\ }\bibinfo {title} {Ultrastrong-coupling circuit
  $\mathrm{QED}$ in the radio-frequency regime},\ \href {\doibase
  10.1103/PhysRevA.100.043815} {\bibfield  {journal} {\bibinfo  {journal}
  {Phys. Rev. A}\ }\textbf {\bibinfo {volume} {100}},\ \bibinfo {pages}
  {043815} (\bibinfo {year} {2019})}\BibitemShut {NoStop}%
\end{thebibliography}%

\end{document}